\documentclass[twocolumn, twocolappendix]{aastex63}

\usepackage[english]{babel}
\usepackage{blindtext}
\usepackage{todonotes}
\usepackage{booktabs}
\usepackage{gensymb}
\usepackage{float}
\usepackage{newtxtext,newtxmath}
\usepackage[T1]{fontenc}
\usepackage{ae,aecompl}
\usepackage{graphicx}	% Including figure files
\usepackage{amsmath}	% Advanced maths commands
\usepackage{amssymb}	% Extra maths symbols
\usepackage{multirow}
\newsavebox{\leftbox}
\newsavebox{\rightbox}
\shorttitle{\textit{NICER} Observations of Three MSPs}
\shortauthors{Rowan et al.}

\begin{document}

\title{A NICER VIEW OF SPECTRAL AND PROFILE EVOLUTION FOR THREE X-RAY EMITTING MILLISECOND PULSARS}

%\correspondingauthor{Andrea Lommen}
\email{dmrowan@haverford.edu, alommen@haverford.edu}

% The list of authors, and the short list which is used in the headers.
% If you need two or more lines of authors, add an extra line using \newauthor

\author[0000-0003-2431-981X]{Dominick M. Rowan}
\affiliation{Haverford College,
370 Lancaster Ave, Haverford PA, 19041, USA}

\author{Zaynab Ghazi}
\affiliation{Haverford College,
370 Lancaster Ave, Haverford PA, 19041, USA}

\author{Lauren Lugo}
\affiliation{Haverford College,
370 Lancaster Ave, Haverford PA, 19041, USA}

\author{Elizabeth Spano}
\affiliation{Haverford College,
370 Lancaster Ave, Haverford PA, 19041, USA}

\author[0000-0003-4137-7536]{Andrea Lommen}
\affiliation{Haverford College,
370 Lancaster Ave, Haverford PA, 19041, USA}

\author[0000-0001-6119-859X]{Alice Harding}
\affiliation{Astrophysics Science Division, 
NASA Goddard Space Flight Center, Greenbelt, MD 20771, USA}

\author[0000-0002-2666-4812]{Christo Venter}
\affiliation{Centre for Space Research, North-West University, Private Bag X6001, Potchefstroom 2520, South Africa}

\author[0000-0002-8961-939X]{Renee Ludlam}
\affiliation{Cahill Center for Astronomy and Astrophysics, 
California Institute of Technology, Pasadena, CA 91125, USA}
\affiliation{Einstein Fellow}

\author[0000-0002-5297-5278]{Paul S. Ray}
\affiliation{U.S. Naval Research Laboratory, Washington, DC 20375, USA}

\author[0000-0002-0893-4073]{Matthew Kerr}
\affiliation{U.S. Naval Research Laboratory, Washington, DC 20375, USA}

\author{Zaven Arzoumanian}
\affiliation{Astrophysics Science Division, 
NASA Goddard Space Flight Center, Greenbelt, MD 20771, USA}

\author[0000-0002-9870-2742]{Slavko Bogdanov}
\affiliation{Columbia Astrophysics Laboratory, Columbia University,
550 West 120th Street, New York, NY, 10027, USA}

\author{Julia Deneva}
\affiliation{George Mason University,
resident at Naval Research Laboratory, Washington, DC 20375, USA}

\author[0000-0002-6449-106X]{Sebastien Guillot}
\affil{IRAP, CNRS, 9 avenue du Colonel Roche, BP 44346, F-31028 Toulouse Cedex 4, France}
\affil{Universit\'{e} de Toulouse, CNES, UPS-OMP, F-31028 Toulouse, France}

\author[0000-0003-0771-6581]{Natalia Lewandowska}
\affiliation{Department of Physics and Astronomy,
West Virginia University, P.O. Box 6315, Morgantown, WV, 26506, USA}
\affiliation{Center for Gravitational Waves and Cosmology
West Virginia University, Chestnut Ridge Research Building, Morgantown, WV, 26506, USA}

\author{Craig B. Markwardt}
\affiliation{Astrophysics Science Division, 
NASA Goddard Space Flight Center, Greenbelt, MD 20771, USA}

\author[0000-0001-5799-9714]{Scott Ransom}
\affiliation{NRAO, 520 Edgemont Rd., Charlottesville, VA 22903, USA}

\author[0000-0003-1244-3100]{Teruaki Enoto}
\affiliation{Department of Astronomy, Kyoto University,
Kitashirakawa-Oiwake-cho, Sakyo-ku, Kyoto 606-8502, Japan}

\author{Kent S. Wood}
\affiliation{Praxis, resident at the Naval Research Laboratory, Washington, DC 20375, USA}

\author{Keith C. Gendreau}
\affiliation{Astrophysics Science Division, 
NASA Goddard Space Flight Center, Greenbelt, MD 20771, USA}

% These dates will be filled out by the publisher
%\date{Accepted 01/28/2020. Received YYY; in original form ZZZ}

% Enter the current year, for the copyright statements etc.

% Don't change these lines
%\hypersetup{draft}

\label{firstpage}
%\maketitle

\renewcommand{\abstractname}{ABSTRACT}
\begin{abstract} 
We present two years of Neutron star Interior Composition Explorer (\textit{NICER}) X-ray observations of three energetic rotation-powered millisecond pulsars (MSPs): PSRs~B1937$+$21, B1821$-$24, and J0218$+$4232. We fit Gaussians and Lorentzians to the pulse profiles for different energy sub-bands of the soft X-ray regime to measure the energy dependence of pulse separation and width. We find that the separation between pulse components of PSR J0218$+$4232 decreases with increasing energy at $>3\sigma$ confidence. The $95\%$ upper limit on pulse separation evolution for PSRs~B1937$+$21 and B1821$-24$ is less than 2 milliperiods per keV. Our phase-resolved spectral results provide updated constraints on the non-thermal X-ray emission of these three pulsars. The photon indices of the modeled X-ray emission spectra for each pulse component of PSR~B1937$+$21 are inconsistent with each other at the 90\% confidence level, suggesting different emission origins for each pulse. We find that the PSR~B1821$-$24 and PSR~J0218$+$4232 emission spectra are invariant with phase at the 90\% confidence level. We describe the implications of our profile and spectral results in the context of equatorial current sheet emission models for these three MSPs with non-thermal, magnetospheric X-ray emission. 

\end{abstract}


%%%%%%%%%%%%%%%%%%%%%%%%%%%%%%%%%%%%%%%%%%%%%%%%%%

%%%%%%%%%%%%%%%%% BODY OF PAPER %%%%%%%%%%%%%%%%%%

\section{Introduction}
Despite the detection of approximately 325 millisecond pulsars (MSPs), no unified emission theory describes the population of these energetic objects. MSPs are a distinct class of rotation-powered pulsars with spin periods $P\lesssim10$ milliseconds and low spin-down rates $\dot{P} \leq 10^{-18}\rm{s}/\rm{s}$. A majority of MSPs have a binary companion \citep{Becker99, Lorimer08}, suggesting that MSPs are `recycled' pulsars spun-up through accretion of matter from companion stars \citep{Bhattacharya91}. This accretion history may add complexity to the already compact MSP magnetosphere; the resulting variety of observed MSP emission properties challenges generalizations of emission theory \citep{Rankin17, Johnson14}.  

Study of non-thermal pulsar emissions can aid in the modeling of magnetosphere structures \citep{Kalapotharakos14}. MSP emission has been studied extensively in the radio regime \citep[e.g.,][]{Kramer98, Eilek16}, often as part of pulsar timing arrays \citep{Hobbs10}.  Comparisons of these MSP radio measurements with higher energy observations have shown that pulse profiles can vary dramatically in number of components and phase separation at different energies \citep{Johnson13}. Observations of high energy emission therefore offer a complementary perspective on study of MSP emission processes.
 
The pulsed high energy emission can either be thermal or non-thermal in origin. Though much of the surface of old MSPs is too cold to significantly emit at high energies, the polar cap model describes how regions at the magnetic poles can emit thermal radiation in the soft X-ray band due to surface heating by particles that move along magnetic field lines towards the cap \citep{Harding01, Cerutti16}. \citet{Riley19} used models of these hot spots to measure the neutron star mass and radius of PSR J0030$+$0451. Some nearby MSPs have had their surface temperatures constrained through observation of polar cap emission \citep[e.g.,][]{Durant12,Rangelov17,Guillot19}. Non-thermal emission likely originates in the outer magnetosphere in gaps of low plasma density, such as the slot gap \citep{Arons83, Muslimov03, Dyks03} or outer gap \citep{Cheng86, Venter14}. In these areas, bounded by the last closed magnetic field line and the null charge surface (in the outer gap case), particles are accelerated by strong electric fields.

Recent models investigating global particle simulations have shifted the focus towards the current sheet (CS) as a source of high energy emission \citep{Kalapotharakos14, Cerutti16}. Gamma-ray pulsar models place the dissipation regions near the separatricies that intersect at the Y-point, where the last closed field line meets the light cylinder (LC), and beyond the LC near the equatorial CS \citep[e.g.,][]{Bambrilla18}. 
While some kinetic models suggest the high-energy gamma-ray emission is primarily due to particles that have been energized by magnetic reconnection in the CS and radiate synchrotron emission up to the GeV band \citep{Philippov18}, others  interpret gamma-ray emission to be curvature radiation, but predict X-ray emission due to synchrotron from both primary and secondary particles \citep{Harding15, Harding18}. In both cases, the emission from these regions maps onto caustics that are probed by observations. 

The observed pulse profile can be understood in the context of skymaps of this emission for a given pulsar obliquity and viewing angle \citep{Kalapothrakos18,Philippov18}. Each peak of the pulse profile occurs when the line of sight passes through the CS. The models of equatorial current sheet and Y-point emission can consistently reproduce the double-peaked gamma-ray profiles \citep[e.g.,][]{Bai10}. 

For pulsars with X-ray pulse profiles nearly in phase with the corresponding gamma-ray profiles, the emission site of the non-thermal X-ray photons must be close to that of the gamma rays \citep{Venter12}. Pulsars with observed multi-wavelength phase-alignment, such as the Crab pulsar \citep{Ansoldi16}, make up a small subset of the high-energy population. The study of pulsars in this select group therefore represents a unique opportunity to relate observed profile and spectral features to emission theories. 

Component separation has been observed to decrease with increasing radio frequency \citep[e.g.,][]{Hankins91}. High-energy emission of the Crab pulsar has been studied extensively \citep[e.g.,][]{Eikenberry96}, and indicates morphology variations in the time and energy domains \citep{Ge16, Mukerjee99, Eikenberry97}. Phase resolved spectral analysis of pulse components and pulse edges has also been conducted on a handful of bright sources \citep[e.g.,][]{Rots98, Pravdo97}. Due to a scarcity of counts, these high-energy measurements have been generally limited to a handful of bright MSPs. X-ray emission of MSPs has been studied with a variety of telescopes, including \textit{ROSAT} \citep{Becker93}, \textit{ASCA} \citep{Takahashi01}, \textit{BeppoSAX} \citep{Nicastro04}, \textit{RXTE} \citep{Cusumano03}, \textit{Chandra} \citep{Zavlin02}, \textit{XMM-Newton} \citep{Ng14}, and \textit{NuSTAR} \citep{Gotthelf17}.

\cite{Deneva19} presented initial Neutron star Interior Composition Explorer (\textit{NICER}) timing results for three relatively bright X-ray MSPs---PSRs~B1821$-$24, B1937$+$21, and J0218$+$4232---demonstrating the precise timing and energy measurement capabilities of the instrument. Here, we present two years of \textit{NICER} X-ray observations between 0.2--12.0~keV of the same three MSPs. We produce phase-folded profiles for different energy sub-bands in soft X-rays and model the phase-resolved emission spectra for each pulse component. We use these results to offer insight into the origin of high energy MSP emission. Section \S\ref{observations_section} describes \textit{NICER} observation parameters and filtering. Sections~\S\ref{1937_section}, \S\ref{1821_section}, and \S\ref{0218_section} describe the pulse profiles and emission spectra of PSR~B1937$+$21, PSR~B1821$-$24, and PSR~J0218$+$4232, respectively. Finally, we summarize our results in Section \S\ref{discussion_conclusion_section}.

\section{NICER Observations} \label{observations_section}

\textit{NICER} \citep{Gendreau17}
was deployed on the \textit{International Space Station} in June 2017. \textit{NICER} observes X-rays between 0.2~keV and 12~keV with a peak collecting area of 1900 cm$^2$ at 1.5~keV. Its X-ray Timing Instrument (XTI) is made up of 56 Focal Plane Modules (FPMs, 52 operational on orbit), each containing a silicon drift detector \citep{Prigozhin16} associated with an X-ray concentrator optic \citep{Okajima16}. The sensitive area of each detector is deliberately small to mitigate particle background and to minimize the electron drift times between the photon interaction sites and the detector anode. 

Each photon detection event is processed by two analog signal processing chains: a ``slow'' chain with a 465 ns peaking time, and a ``fast'' chain with an 84 ns peaking time \citep{Prigozhin16}. With greater noise reduction due to the slower rise time, the slow chain offers more-accurate pulse height measurements; the fast chain, optimized for timing, is noisier and does not reliably trigger for photon energies below $\sim 0.5$~keV. The instrument's overall photon time-stamping precision is better than 100 ns rms, traceable to UTC via an onboard GPS system. 

Data were processed with {\tt HEASOFT} v.6.26.1  and the \textit{NICER} specific {\tt NICERDAS} v.5, with Calibration Database ({\tt CALDB}) version 20190520. We apply the following standard criteria in data reduction:
\begin{itemize}
    \item Pointing offset is $<0.015\degree$ from source
    \item Elevation above Earth limb is $>20\degree$,  increased to $>30\degree$ in the case of bright Earth. 

\end{itemize}
Additional filters are set based on detector overshoot (indicative of high radiation backgrounds) and undershoot (indicative of high optical loading on the detectors) rates:
\begin{itemize}
    \item {\tt FPM\textunderscore OVERONLY\textunderscore COUNT} $ < 1$
    \item { \tt FPM\textunderscore OVERONLY\textunderscore COUNT} $ < 1.52\times{\tt COR}$\textunderscore {\tt SAX}$^{-0.633}$
    \item {\tt FPM\textunderscore UNDERONLY\textunderscore COUNT} $<200$,
\end{itemize}
where these parameters are defined and derived by the mission's data-processing pipeline. We include additional criteria on the cutoff rigidity {\tt COR\_SAX} (in units of GeV/$c$) using the geomagnetic activity index, $K_p$ \citep{Bartels39}:
\begin{itemize}
    \item $K_p$ $<5$
    \item {\tt COR\textunderscore SAX } $> 1.914\times K_p^{0.684}+0.25$.
\end{itemize}
Finally, the ratio of pulse-height amplitudes measured in each processing chain, {\tt PI\textunderscore RATIO}, is a useful diagnostic for filtering out background events during data processing. We attempt more restrictive cuts than the default used in {\tt nicerclean}, but find no significant variation in modeling of our pulse profiles or emission spectra. We therefore apply the default cut to {\tt PI\textunderscore RATIO},
\begin{equation}
    {\tt PI\_RATIO} \equiv {\tt PI/PI\_FAST} > 1.1 + 120/{\tt PI}.
\end{equation}

\section{PSR~B1937$+$21} \label{1937_section}
PSR~B1937$+$21 (also known as PSR~J1939$+$2134) was the first MSP, discovered by \cite{Backer82}. Later, the second pulse component was detected by \cite{Nicastro04}. X-ray observations reveal  nearly a 100\% pulsed fraction \citep{Ng14}. PSR~B1937$+$21 has a spin period $P=1.558$~ms and a spin down luminosity of $\Dot{E} = 1.1 \times 10^{36}$~erg\,s$^{-1}$. X-ray emission of this source has been studied with \textit{ASCA} \citep{Takahashi01}, \textit{BeppoSAX} \citep{Nicastro04}, \textit{RXTE} \citep{Cusumano03, Guillemot12}, \textit{NuSTAR} \citep{Gotthelf17}, \textit{Chandra} \citep{Zavlin07}, and \textit{XMM-Newton} \citep{Ng14}. \textit{NICER} observed this source for $\sim1340$ ks between 2017 June 28 and 2019 June 24, in 379 ObsIDs.  

\subsection{Pulse Profiles} \label{1937_profile_section}

\begin{table*}
    \begin{lrbox}{\leftbox}
    \begin{tabular}{cc}
\toprule
Energy range (keV) & P1/P2\\
\midrule
0.5--2.0 & 8.2$\pm$ 1.7\\
2.0--4.0 & 7.7$\pm$ 0.7\\
4.0--12.0 & 7.1$\pm$1.1\\
\bottomrule
\end{tabular}
%updated with new phase data
    \end{lrbox}

    \begin{lrbox}{\rightbox}
    \begin{tabular}{llcccrr}
\toprule
           Region &   Phase Range &       $\Gamma$ &  $\chi^2_{\nu}$ &  Degrees of &  Number of \\
& & & & Freedom & Photons\\ 
\midrule
 \textbf{Pulse 1} &   0.03 -- 0.1 &  0.98$\pm$0.04 &            1.02 &       101.0 &      61292 \\
                  &  0.04 -- 0.09 &  0.99$\pm$0.04 &            0.95 &        85.0 &      48069 \\
     Leading Edge &  0.03 -- 0.05 &   0.94$\pm$0.1 &            0.74 &        55.0 &      15871 \\
                  &  0.04 -- 0.05 &  0.91$\pm$0.09 &            0.65 &        34.0 &       9118 \\
    Trailing Edge &  0.05 -- 0.15 &  0.97$\pm$0.06 &            0.88 &       117.0 &      75564 \\
                  &  0.05 -- 0.14 &  0.96$\pm$0.05 &            0.81 &       107.0 &      69658 \\
\midrule
 \textbf{Pulse 2} &  0.56 -- 0.61 &    0.6$\pm$0.2 &            1.05 &        62.0 &      32590 \\
                  &   0.57 -- 0.6 &    0.6$\pm$0.2 &            1.19 &        39.0 &      20108 \\
\bottomrule
\end{tabular}

    \end{lrbox}

    \centering
    \makebox[0pt]{%
        \begin{minipage}[b]{\wd\leftbox} % A minipage that covers half the page 
            \centering
            \caption{Evolution of PSR~B1937$+$21 P1/P2 peak ratio with respect to energy.} \label{tab:1937_PR}
            \usebox{\leftbox}
        \end{minipage}\quad
        \begin{minipage}[b]{\wd\rightbox}
            \centering
            \caption{Best-fit absorbed single power-law model parameters for PSR~B1937$+$21 emission spectra at different phase selections. Since the number of counts vary with each phase selection, different binning is used. This is reflected in the DOF used in the fit. For each fit we freeze the absorbing column density $N_{\rm{H}}=1.87\times 10^{22}$~cm$^{-2}$, the value fit by the model with both components.} \label{tab:1937_models}
            \usebox{\rightbox}
        \end{minipage}%
    }
\end{table*}

\begin{figure}
    \centering
    \includegraphics[width=\linewidth]{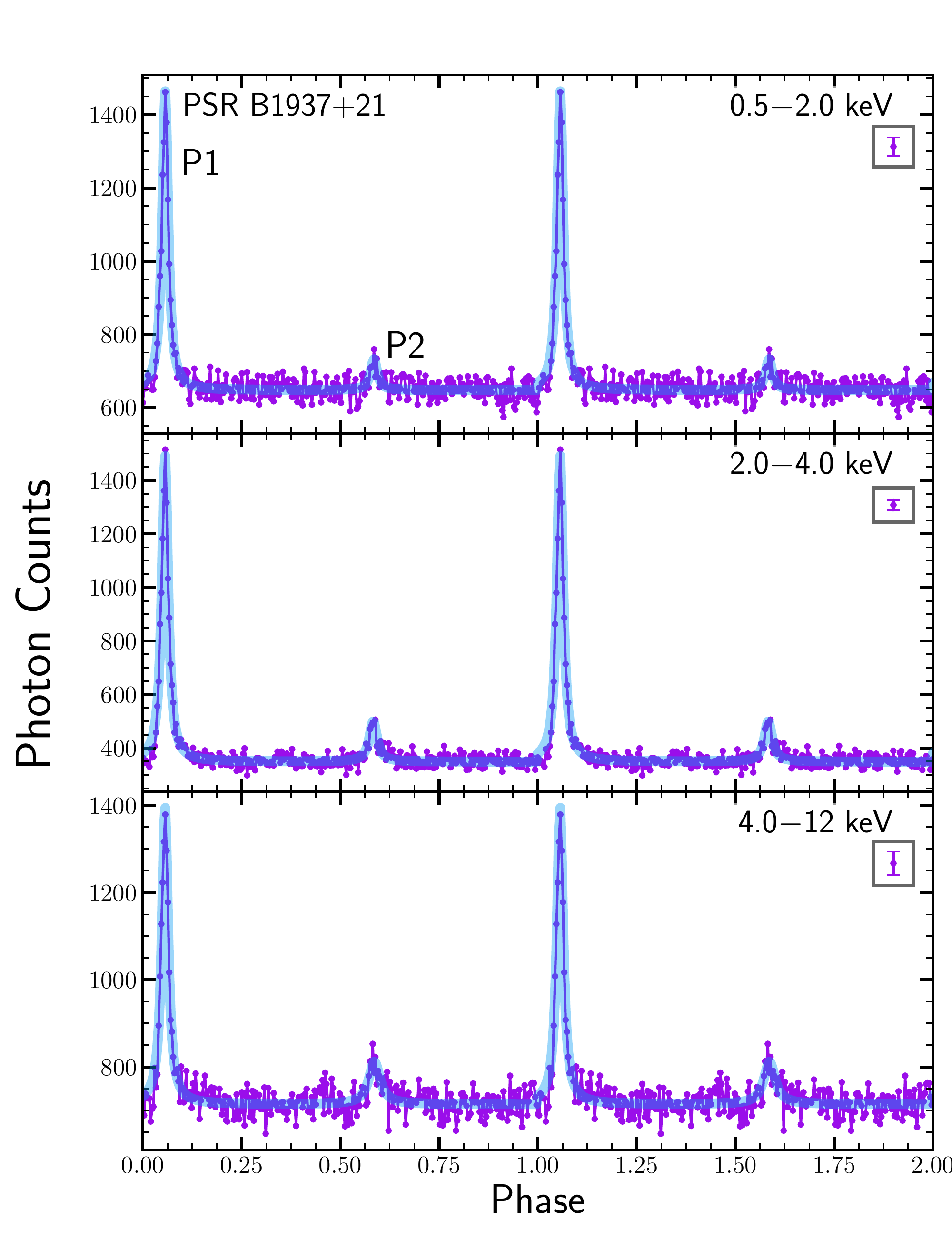}
    \caption{Phase folded pulse profiles for PSR~B1937$+$21 at three different energy selections with 300 phase bins. The best fit two Lorentzian model is plotted in blue. The boxed point shows the characteristic error bar for each profile. }
    \label{fig:1937_profiles}
\end{figure}

Figure~\ref{fig:1937_profiles} shows the phase-folded profiles over three energy ranges encompassing the entire \textit{NICER} passband. For all pulsars, we use the same timing model as \citet{Deneva19} where the radio pulses are at phase 0. We observe two pulse components, which we label as P1 and P2.   

\begin{figure}
    \centering
    \includegraphics[width=\linewidth]{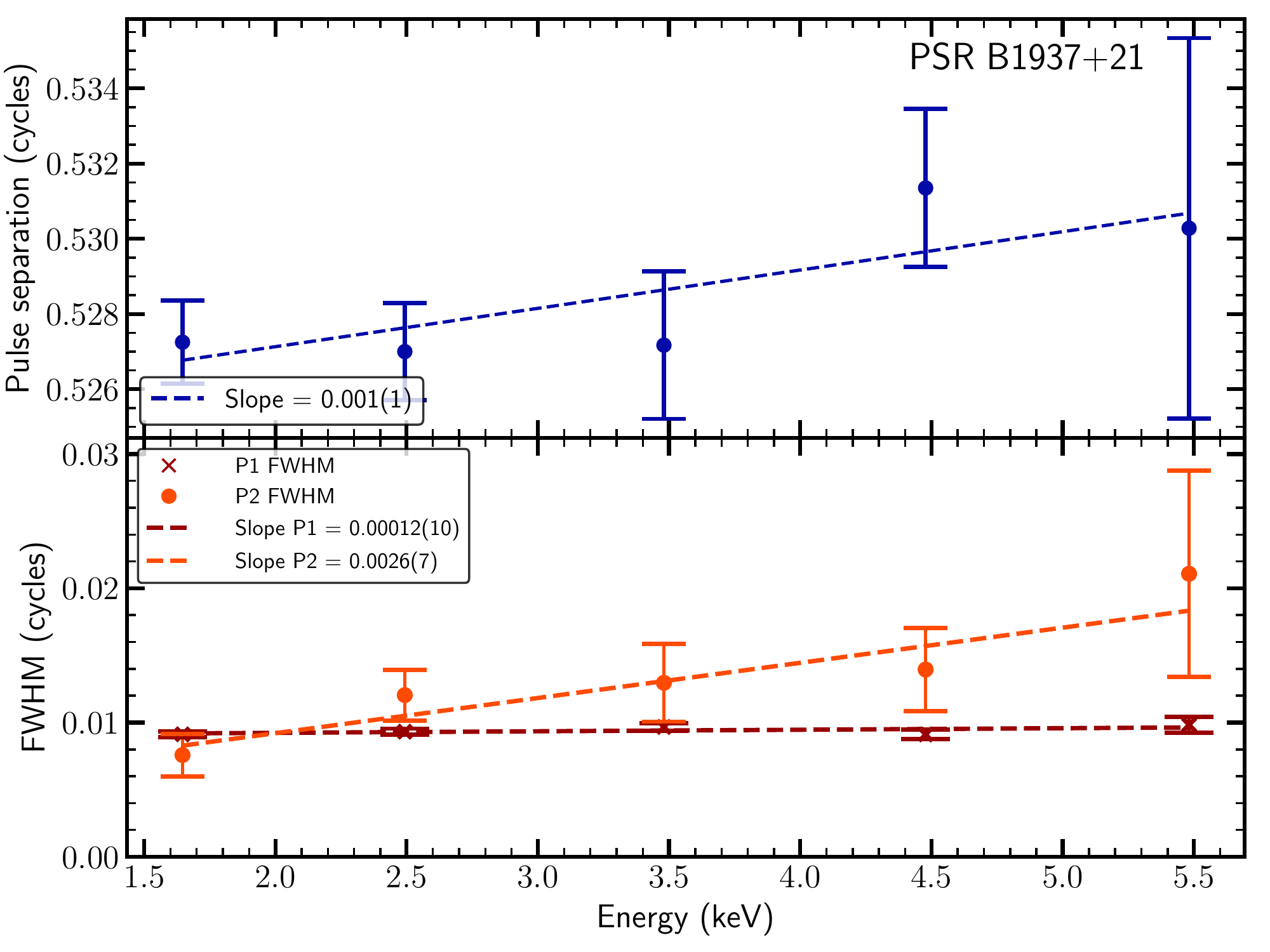}
    \caption{Evolution of pulse separation and FWHM as a function of energy for PSR~B1937$+$21. A weighted linear regression is used to fit this evolution over five 1~keV-wide energy intervals whose corresponding data points were located through spectra integration.}
    \label{fig:1937_PP}
\end{figure}

We apply a three-step fitting procedure, first to extract the parameters of the peaks such as position and width, and second to measure the change in those parameters as a function of energy. In the first step, for profiles in each energy bin spanning 1~keV, we fit a Lorentzian to each pulse component to measure the positions, widths and amplitudes. We provide a comparison of profile fits with different functional forms in Appendix \ref{appendixFitting}. We then calculate the median energy for the energy bin using the modeled pulsed emission spectra described below in Section \S\ref{1937_Spectro}. In the third step, we perform weighted linear fits to measure the slopes of the pulse separation, $m_{\rm{sep}}$, and FWHM of each peak, $m_{\rm{FWHM,P1}}$ and $m_{\rm{FWHM,P2}}$, as a function of energy (Figure ~\ref{fig:1937_PP}), with the weights being the 1$\sigma$ standard deviation errors on the measurement of each profile feature. If the pulse profile is consistent across the energy range, we expect the slopes to be zero. 

We assess the significance of each slope using a $\Delta\chi^2$ test to compare the best-fit slope with the null hypothesis of a zero slope. We find that $m_{\rm{sep}}$ is consistent with zero for PSR~B1937$+$21. Using the $\chi^2$ distribution we find that the $95\%$ upper limit for $m_{\rm{sep}}$ is 0.0019 cycles/keV. The separation is defined as the distance in phase between P1 and P2, specifically the space after P1 and before P2 as labeled on Figure ~\ref{fig:1937_PP}. This translates to the absolute value $|P2 - P1|$, where $P1$ and $P2$ are the centers of the double-peaked Lorentzian fit. 

\textit{NICER} observations show that for this pulsar, there is a marginal increase in pulse width at the 1.0$\sigma$ confidence level for P1, and 2.6$\sigma$ for P2. We find that the $95\%$ upper limit of $m_{\rm{FWHM,P1}}$ is 0.00021 cycles/keV. The slope $m_{\rm{FWHM,P2}}$ is an order of magnitude higher with 0.0033 cycles/keV.

Table \ref{tab:1937_PR} shows the evolution of the peak-ratio of these pulses with respect to energy. We calculate the peak ratio and corresponding error using the amplitude of the Lorentzian fits with the background offset taken into account. We find that the ratio of the pulse heights P1/P2, decreases with increasing energy which suggests that the spectral behaviors of the two peaks are dissimilar.

\setlength{\tabcolsep}{6pt}
\begin{table}
\centering
    \caption{Pulse separation and width corresponding to Figure \ref{fig:1937_PP}. For each energy range, we use the pulsed emission spectra from \S\ref{1937_Spectro} to find the median energy. We find the slope of the pulse separation is consistent with zero.}
    \label{tab:1937_PP}
    \begin{tabular}{cccccc}
\toprule
 Pulse & FWHM & FWHM & Energy & Energy \\
 Separation & of P1 &of P2 & band&Median\\
  (cycles)&(cycles)&(cycles)&(keV)&(keV)\\
\midrule
0.527(1)  & 0.0091(2) & 0.008(1) & $1.0-2.0$ & $1.65$\\
0.527(1) & 0.0093(2)& 0.012(2) & $2.0-3.0$ &  $2.49$ \\
0.527(2) & 0.0097(3) &  0.013(3)  & $3.0-4.0$ & $3.48$\\
0.531(2) & 0.0091(4) & 0.014(3) &  $4.0-5.0$ & $4.48$\\
0.530(5) & 0.0098(6) & 0.021(8)  & $5.0-6.0$ & $5.48$\\
\bottomrule
\end{tabular}

\end{table}
\subsection{Spectroscopy} \label{1937_Spectro}
We extract the source spectrum with {\tt XSELECT} v.2.4 from the cleaned event files used in the previous section. All spectral analyses use the \textit{NICER} redistribution matrix file and ancillary response file versions 1.02. Since the pulsed fraction is nearly 100\% for this source \citep{Ng14} we assume that any detected photons within the off-pulse phase range are due to the X-ray background. We define the off-pulse region to be $0.20 \leq \phi \leq 0.40$, and $0.7 \leq \phi \leq 1.0$. We use the spectrum extracted from these phase ranges as the background spectrum in our subsequent spectral analysis. 

We select the on-pulse region using the variance of the background emission. Using 100 phase bins, the phase limits of each component are chosen to be where the counts exceed 3 times the standard deviation of the background. The P1 phase region is thus $0.03 \leq \phi \leq 0.10$ and the P2 phase region is $0.56 \leq \phi \leq 0.61$. 

We compute the difference spectra by subtracting the spectrum of the background from the spectrum of the pulsed components \citep[e.g.,][]{Fabian03}. Since the phase range is wider for the off-pulse emission spectra, we re-scale the spectrum by adjusting the exposure time with the ratio of the two phase widths. We fit models to the spectra with {\tt XSPEC} v.12.10.1 \citep{Arnaud96}. Interstellar medium (ISM) absorption is set using abundances from \cite{Wilms00} and the {\tt tbabs} model. This model sets the cross-section to Vern cross-sections from \cite{Verner96} and normalizes the total photoionization cross-section by the column density of hydrogen, $N_{\textrm{H}}$. The observed spectrum is given in Equation 2 of \cite{Wilms00} as
\begin{equation}
    I_{\textrm{obs}}(E)=e^{-\sigma_{\textrm{ISM}}(E)N_{\textrm{H}}}I_{\textrm{source}}(E)
\end{equation}
where $\sigma_{\textrm{ISM}}$ is the energy dependent photoionization cross-section of the ISM, $N_{\textrm{H}}$ is the total hydrogen number density, and $I_{\textrm{source}}(E)$ is the intrinsic X-ray spectrum of the source. 

First we model the emission spectra for the two components together, binned with 600 counts per channel. The best-fit single absorbed power-law model for the pulsed emission gives an absorbing column $N_{\textrm{H}} = (1.88 \pm 0.18) \times 10^{22}$~cm$^{-2}$ and spectral index $\Gamma = 0.93 \pm 0.09$ with $\chi^{2}_{\nu} = 0.85$ for 98 degrees of freedom (dof).

As done in previous work \citep[e.g.,][]{Ng14}, we also consider an extra blackbody component, {\tt (powerlaw+bbody)*tbabs}. The fit results in a very high temperature $kT\sim1.1$~keV with $\chi^2_{\nu}=0.82$ for 96 dof. This temperature is almost certainly too large to represent a physical thermal component. \textit{NICER} observations of thermally-emitting MSPs have fit blackbody temperatures at $\sim0.1$~keV \citep[e.g.,][]{Ray19, Harding19}. We can obtain a reasonable fit by constraining the thermal parameter to $kT\leq 0.5$~keV and the photon index in the range $0.8<\Gamma<1.5$ (resulting in $\chi^{2}_{\nu} = 0.99$ with dof$=$96). The blackbody fit with constrained parameters is slightly worse than the absorbed power-law alone. Since a blackbody component is not statistically required by the data, our results are in agreement with previous work and suggest the X-ray spectrum is dominated by nonthermal, magnetospheric emission. 

With \textit{NICER}'s large collecting area and timing uncertainty better than 100~ns, we can model the emission spectrum of each pulse component separately. Similar to measurement of pulse amplitude at different energies, comparison of the modeled spectra between peaks can indicate how the origin of each peak differs. This can help distinguish blended features in pulse components, as done with radio observations of PSR~B1133$+$16 \citep{Hankins91}. We also model the emission spectra between the leading and trailing edges of the primary pulse component. The comparison of pulse edge spectra has been done for bright sources, such as the Crab Pulsar \citep{Eikenberry96} and PSR~B1509$-$58 \citep{Rots98}, but this is the first time this analysis has been possible for PSR~B1937$+$21. 

\begin{figure}
    \centering
    \includegraphics[width=\linewidth]{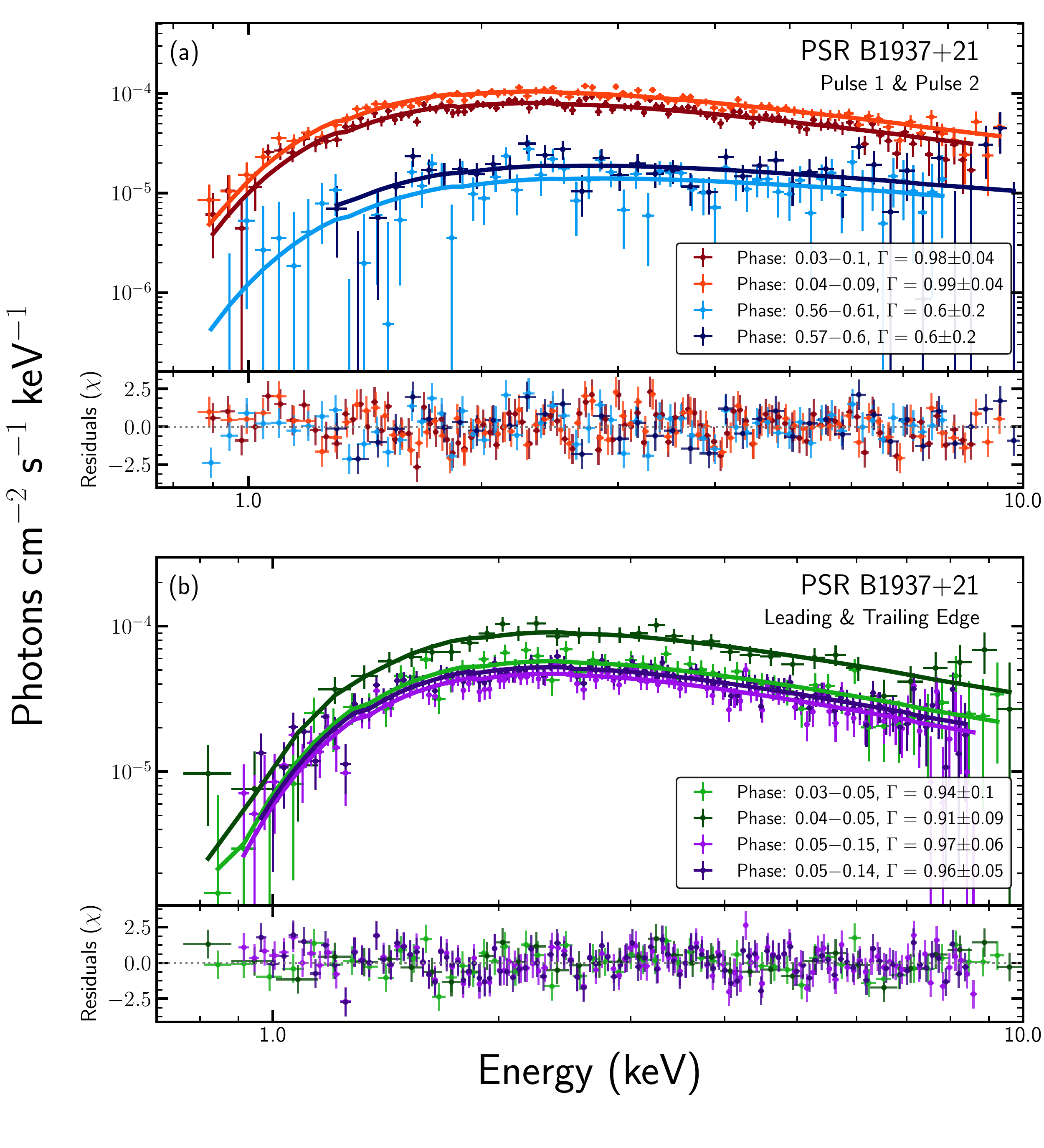}
    \caption{
    Emission spectra of two PSR~B1937+21 pulse components (a) and leading and trailing edges of P1 (b). We fit single absorbed power-law models to each phase selection. Below each set of spectra are the residuals of the best fit powerlaw in units of $\chi$.}
    \label{fig:1937_spectra}
\end{figure}

For each phase selection, we fit a single absorbed power-law. For these fits, we freeze the column density at $N_{\rm{H}}=1.88\times 10^{22}$~cm$^{-2}$, the best fit value from our spectrum of pulsed emission above. Figure~\ref{fig:1937_spectra} shows the emission spectra for a variety of phase selections. For each profile region (P1, leading edge, etc.), we fit multiple phases to compare the emission spectra over narrow variations in phase. Since we find no statistically significant variation between narrow-phases in each region, we make broader comparisons between the two pulses and the leading/trailing edges of P1. The model parameters for each phase-resolved spectra are given in Table \ref{tab:1937_models}. All errors are quoted at the 90\% confidence level.

The difference between the photon indices of the two pulses is statistically significant at the $>90\%$ confidence level. This may suggest that the underlying particle spectrum differs between the two peaks. We find that the 90\% confidence intervals of the photon indices overlap for the leading and and trailing edges. Therefore the \textit{NICER} modeled emission spectra does not differ significantly between the edges of the P1, suggesting a uniform emission origin for the entirety of P1.

\begin{figure}
    \centering
    \includegraphics[width=\linewidth]{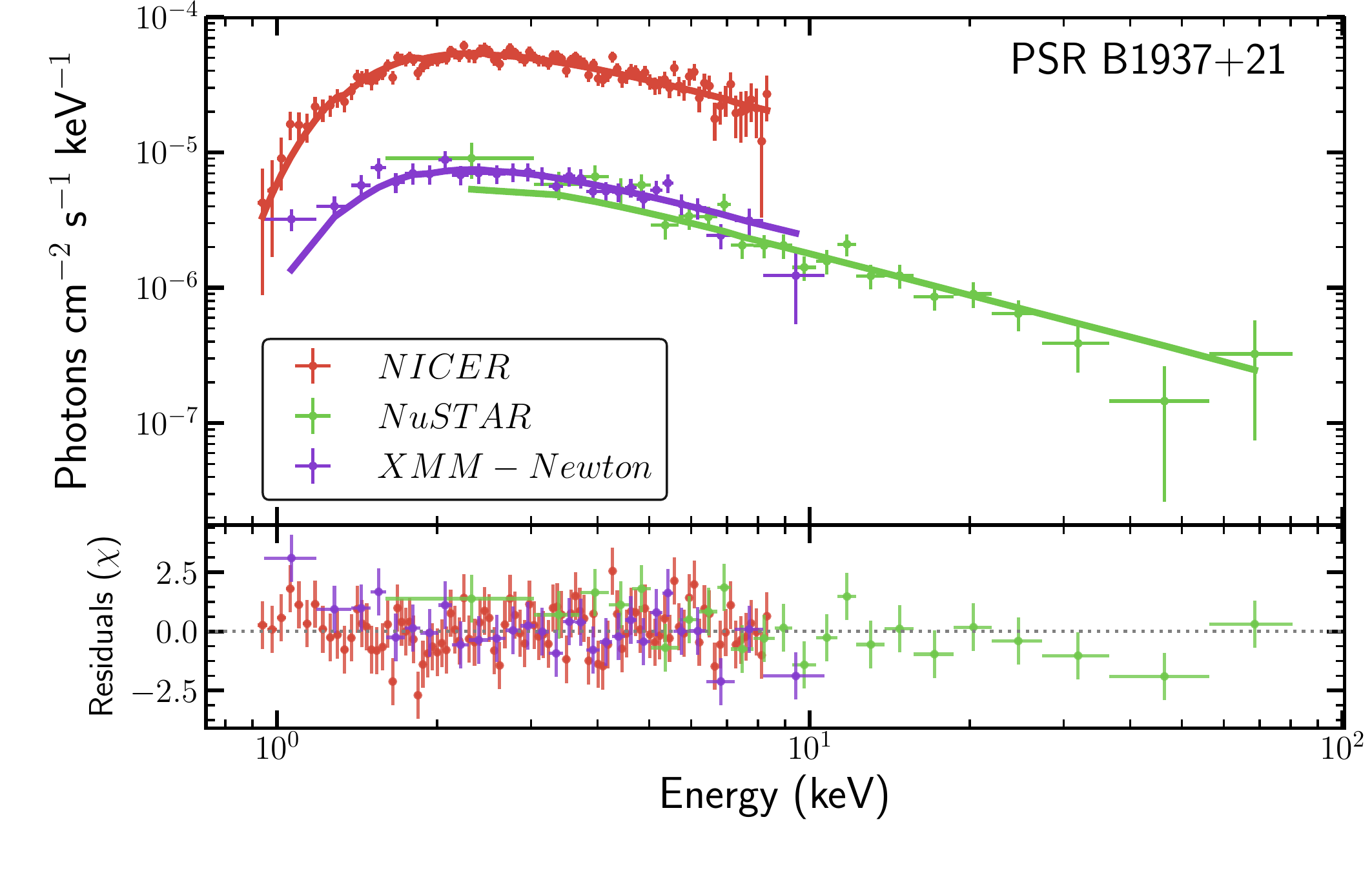}
    \caption{Top: Emission spectra of PSR~B1937+21 primary pulse component fit jointly with data from \textit{NuSTAR} and \textit{XMM-Newton EPIC MOS}. Bottom: Residuals of the best fit power-law model.}
    \label{fig:1937_joint_spectra}
\end{figure}

We use the extracted \textit{NuSTAR} and \textit{XMM-Newton} spectra from \cite{Gotthelf17} to model the broadband emission spectrum of PSR~B1937$+$21. The \textit{NuSTAR} spectrum is also a difference spectra, using the off-pulse emission as the background. The \textit{XMM-Newton} data was originally presented in \cite{Ng14} and uses nearby chip regions to extract a background spectra. \cite{Gotthelf17} previously reported $N_{\textrm{H}}=(1.8\pm0.3)\times10^{22}$~cm$^{-2}$ and spectral index $\Gamma = 1.16 \pm 0.11$ with a $\chi^2_{\nu} = 0.91$ for 83 dof. We use the extracted spectra from \cite{Gotthelf17} to produce a simultaneous fit of \textit{NuSTAR}, \textit{XMM-Newton}, and \textit{NICER}. Since we are fitting difference spectra of pulse emission, the spectra from each telescope differ by multiplicative constants. Rather than fit for these constants, we allow the normalization for each data set to remain a free parameter. We find the best-fit single absorbed power-law has a photon index $\Gamma = 1.04 \pm 0.08$ and a hydrogen absorbing column density $N_{\textrm{H}} = (2.00\pm0.16)\times 10^{22}$~cm$^{-2}$ with a $\chi^2_{\nu} = 0.95$ for 145 dof. Figure \ref{fig:1937_joint_spectra} plots the simultaneous fit, showing the \textit{NICER} spectra index is indeed consistent with measurements made by other telescopes. 

\section{PSR~B1821$-$24} \label{1821_section}

PSR~B1821$-$24 (also known as PSR~J1824$-$2452A) has a 3.05~ms period and was the first radio MSP found in a globular cluster \citep{Lyne87}. Since then, over 150 MSPs have been detected in globular clusters\footnote{\url{http://www.naic.edu/~pfreire/GCpsr.html}}. \textit{ASCA} first detected the X-ray emission from this pulsar \citep{Saito97}, with a spin-down luminosity of $\Dot{E} = 2.2 \times 10^{36}$~erg\,s$^{-1}$ and a X-ray luminosity of $L = 1.3 \times 10^{33}$~erg\,s$^{-1}$. \textit{Chandra} \citep{Becker03,Bogdanov11}, \textit{RXTE}, and \textit{NuSTAR} \citep{Gotthelf17} have observed X-ray emission of PSR~B1821$-$24. No energy dependent phase separation was observed. These studies concluded that the best fit X-ray spectral model of PSR~B1821$-$24 is an absorbed power-law, indicating non-thermal emission. \cite{Gotthelf17} performed phase-resolved spectroscopy to compare the two pulses and found them to have nearly identical photon indices. \textit{NICER} observed PSR~B1821$-$24 for $~\sim715$~ks seconds between 2017-06-25 and 2019-06-30 in 271 ObsIDs.

\subsection{Pulse Profiles} \label{1821_profile_section}

\begin{figure}
    \centering
    \includegraphics[width=\linewidth]{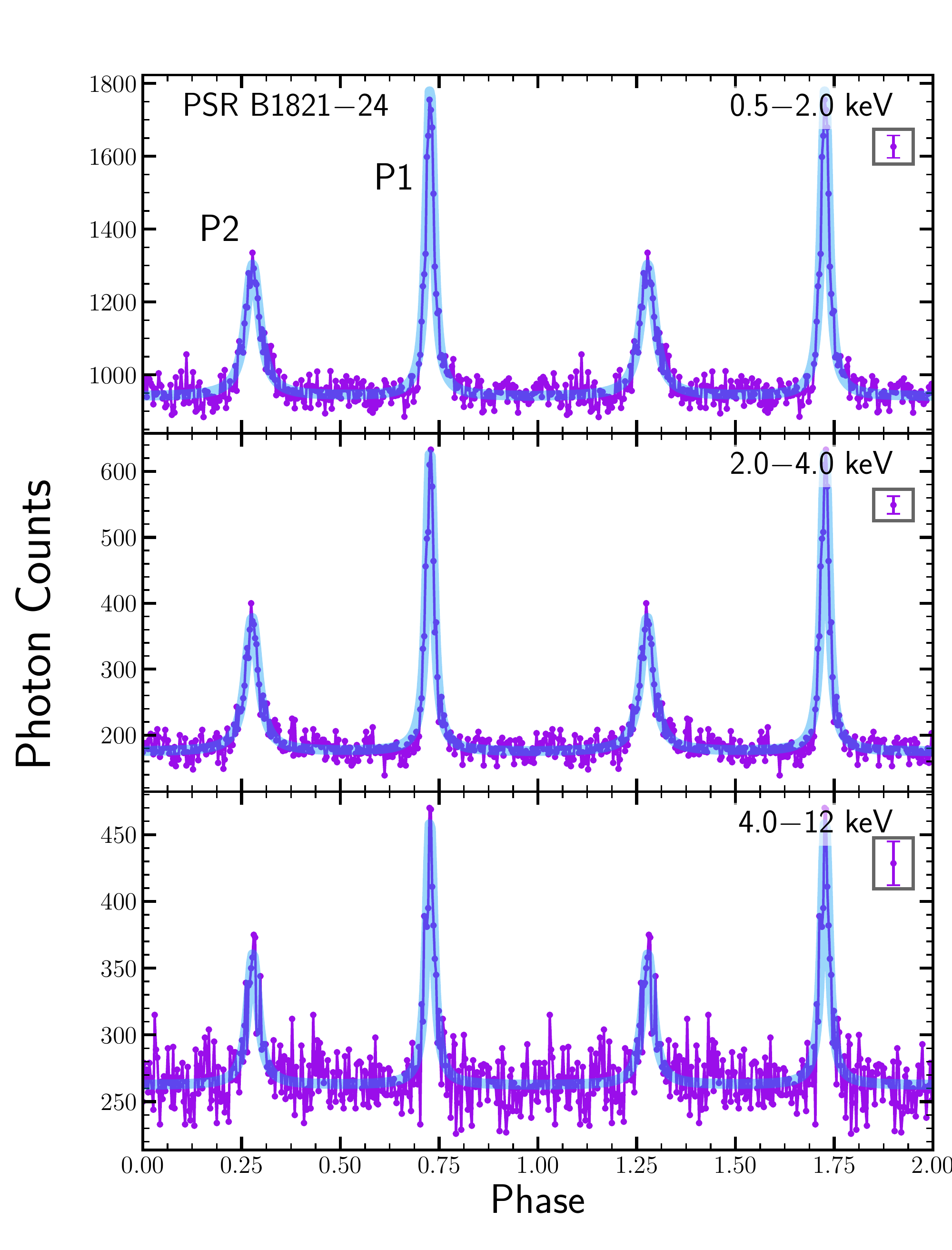}
    \caption{Phase folded pulse profile for PSR~B1821$-$24 with 300 phase bins. We fit a Lorentzian to each pulse to estimate its center and width. P2 in this profile is relatively larger than the secondary pulse for PSR~B1937$+$21. Boxed point shows characteristic error bar for each profile. }
    \label{fig:1821_profiles}
\end{figure}

\begin{table*}
    \begin{lrbox}{\leftbox}
    \begin{tabular}{cc}
\toprule
Energy range (keV) & P1/P2\\
\midrule
0.5--2.0 & 2.3$\pm$0.1\\
2.0--4.0 & 2.3$\pm$0.1\\
4.0--12.0 & 2.0$\pm$0.2\\
\bottomrule
\end{tabular}
    \end{lrbox}

    \begin{lrbox}{\rightbox}
    \begin{tabular}{llcccrr}
\toprule
           Region &   Phase Range &       $\Gamma$ &  $\chi^2_{\nu}$ &  Degrees of &  Number of \\
& & & & Freedom & Photons\\ 
\midrule
 \textbf{Pulse 1} &   0.7 -- 0.78 &  1.12$\pm$0.05 &      1.11 &        55.0 &      49904 \\
                  &  0.71 -- 0.77 &  1.09$\pm$0.05 &      1.04 &        45.0 &      39679 \\
     Leading Edge &   0.7 -- 0.72 &  1.05$\pm$0.12 &      0.77 &        53.0 &      11995 \\
                  &  0.71 -- 0.72 &  0.97$\pm$0.11 &      0.94 &        45.0 &       6651 \\
    Trailing Edge &  0.72 -- 0.78 &  1.13$\pm$0.06 &      0.99 &        64.0 &      37909 \\
                  &  0.72 -- 0.77 &  1.11$\pm$0.05 &      1.29 &        62.0 &      33028 \\
\midrule
 \textbf{Pulse 2} &  0.23 -- 0.34 &  1.14$\pm$0.08 &      1.02 &        45.0 &      60184 \\
                  &  0.24 -- 0.33 &  1.16$\pm$0.08 &      1.09 &        46.0 &      50489 \\
\bottomrule
\end{tabular}

    \end{lrbox}

    \centering
    \makebox[0pt]{%
        \begin{minipage}[b]{\wd\leftbox} % A minipage that covers half the page 
            \centering
            \caption{Evolution of PSR~B1821$-$24 P1/P2 peak ratio with respect to energy.} \label{tab:1821_PR}
            \usebox{\leftbox}
        \end{minipage}\quad
        \begin{minipage}[b]{\wd\rightbox}
            \centering
            \caption{Best-fit model parameters of the absorbed single power-law for PSR~B1821$-$24 emission spectra at different phase selections. Since the number of photons varies between phase selections, we use different binning when extracting the spectra, resulting in different degrees of freedom. The confidence intervals on the photon index are given at the 90\% level.} \label{tab:1821_models}
            \usebox{\rightbox}
        \end{minipage}%
    }
\end{table*}
Figure~\ref{fig:1821_profiles} plots the phase folded profiles for PSR~B1821$-$24 over the three energy ranges spanning the entire \textit{NICER} bandwidth. Like PSR~B1937$+$21, \textit{RXTE} X-ray observations show the pulsed fraction of PSR~B1821$-$24 is near 100\% \citep{Ray08}.

\begin{figure}
    \centering
    \includegraphics[width=\linewidth]{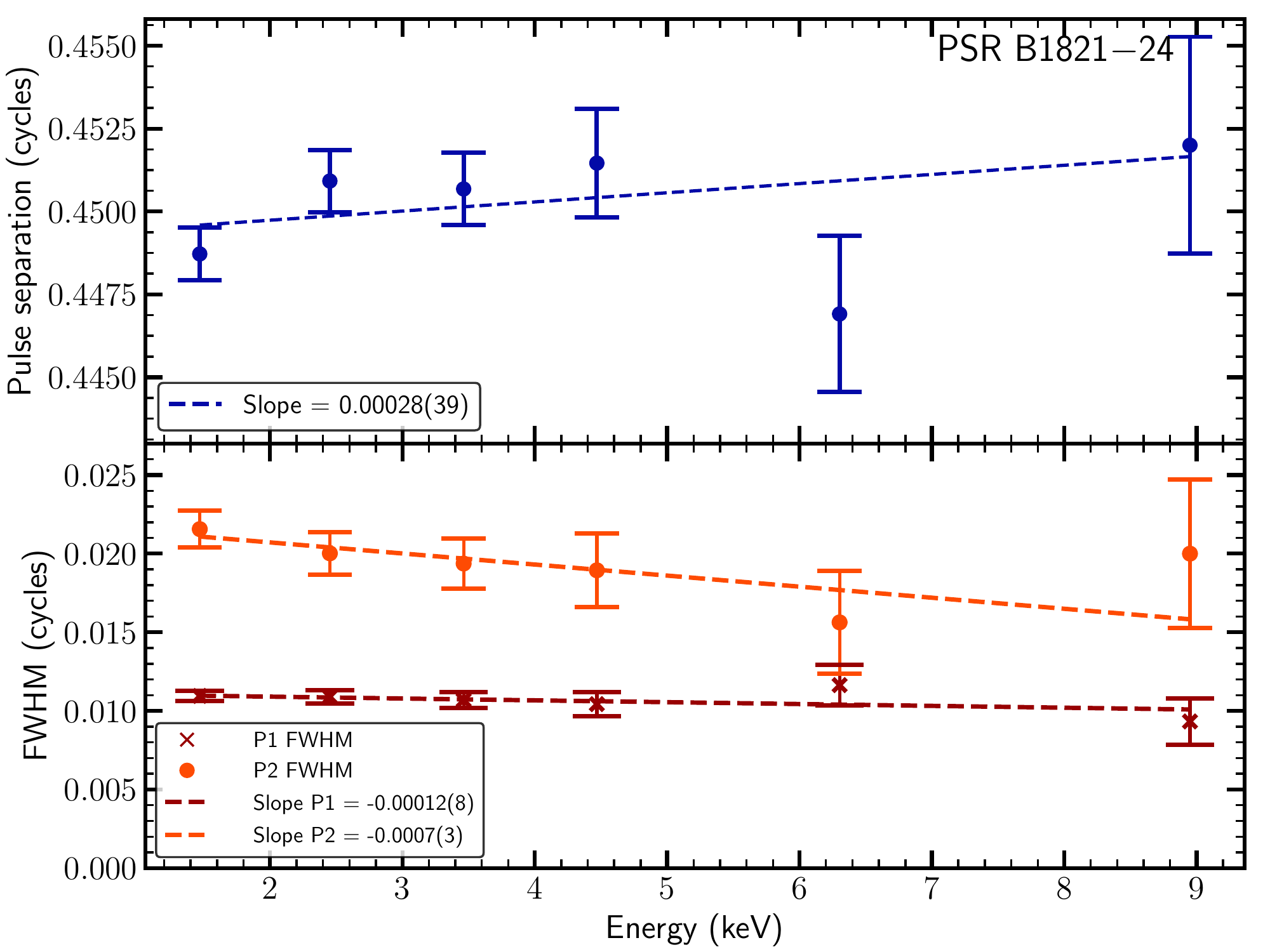}
     \footnotesize{\emph{\newline * The data-point corresponding to the \textit{NICER} range $5.0-8.0$ is at $6.31$ keV through integration of the spectra. This same technique was used to locate the last data point corresponding to \textit{RXTE} data at $8.95$ keV.}}
    \caption{Evolution of pulse separation and FWHM as a function of energy for PSR~B1821$-$24.}
    \label{fig:1821_PP}
   
\end{figure}

PSR~B1821$-$24 has a similar X-ray pulse profile to that of PSR~B1937$+$21 with two narrow pulse components. We apply the same three-step fitting procedure described in Section \S\ref{fig:1937_profiles} by fitting Lorentzians to pulse profiles with narrow energy selections.  Figure~\ref{fig:1821_PP} shows that the slope of the separation as a function of energy, $m_{sep}$ is consistent with zero in the \textit{NICER} energy range. We include an additional data point for the \textit{RXTE} data between $6.0$--$17.0$~keV. The x-axis value for this bin is chosen using the extracted \textit{RXTE} spectrum from \citep{Gotthelf17}. We find the slope $m_{\rm{sep}}$ is again consistent with zero. The $95\%$ upper limit on the pulse separation of PSR~B1821$-$24 is 0.00052 cycles/keV. The \textit{NICER} observations are therefore a testament to the stability of the pulse profile over almost an order of magnitude of X-ray energies. 

The width of both pulses is shown to decrease at the 0.8$\sigma$ confidence level for P1 and 1.5$\sigma$ for P2. The $95\%$ upper limits are 0.00022 and 0.0011, for |$m_{\rm{FWHM,P1}}$| and |$m_{\rm{FWHM,P2}}$|, respectively. Finally, we find that the ratio of the peak components P1/P2, given in Table \ref{tab:1821_PR}, is suggestive of a decrease with increasing energy as was the case with PSR~B1937$+$21. 

\setlength{\tabcolsep}{6pt}
\begin{table}
\centering
    \caption{Pulse separation and width corresponding to Figure \ref{fig:1821_PP}. For each energy range, we use the pulsed emission spectra from \S\ref{1821_spectra_section} to find the median energy. We find the slope of the pulse separation is consistent with zero. The last energy range corresponds to \textit{RXTE} data.}
    \label{tab:1821_PP}
    \begin{tabular}{cccccc}
\toprule
 Pulse & FWHM & FWHM & Energy & Energy \\
 Separation & of P1 &of P2 & band&Median\\
  (cycles)&(cycles)&(cycles)&(keV)&(keV)\\
\midrule
0.449(1)  & 0.0109(3) & 0.021(1)& $1.0-2.0$ & $1.47$\\
0.451(1) & 0.0109(4)& 0.020(1) & $2.0-3.0$ &  $2.45$ \\
0.451(1)&0.0107(5)&  0.019(1) & $3.0-4.0$ & $3.46$\\
0.451(2) & 0.0104(8) & 0.019(2) &  $4.0-5.0$ & $4.47$\\
0.446(2) & 0.0116(10) & 0.016(3) & $5.0-8.0$ & $6.30$\\
0.452(3) &0.0093(10) & 0.020(5) & $6.0-17.0$ & $8.95$\\
\bottomrule
\end{tabular}
\end{table}

\subsection{Spectroscopy} \label{1821_spectra_section}
Following the same procedure described in \ref{1937_Spectro}, we first extract the spectrum of the pulsed emission with a minimum of 600 counts per channel. We define the off-pulse regions to be $0.38 \leq \phi \leq 0.6$ and $0.9 \leq \phi \leq 1.2$. Using the standard deviation of this off-pulse region in a profile with 100 phase bins, we select the P1 phase region as $0.70 \leq \phi \leq 0.78$ and the P2 region as $0.23 \leq \phi \leq 0.34$. The best-fit model of the observed \textit{NICER} emission spectrum extracted from both P1 and P2 is a single absorbed power-law with an absorbing column $(0.40 \pm 0.08) \times 10^{22}$~cm$^{-2}$ and spectral index $\Gamma=1.12\pm0.08$ with $\chi^2_{\nu} = 0.93$ for 101 dof. 

We add a blackbody component to the absorbed power-law and find a $kT$ value too high to suggest the presence of a physical thermal component ($\sim1$~keV). When the temperature is constrained such that $kT\leq0.5$~keV, both the $kT$ and component normalization are consistent with zero within the $90\%$ confidence intervals. Therefore, the PSR~B1821$-$24 spectrum is dominated by magnetospheric emission and the best fit model is a single absorbed power-law.

\begin{figure}
    \centering
    \includegraphics[width=\linewidth]{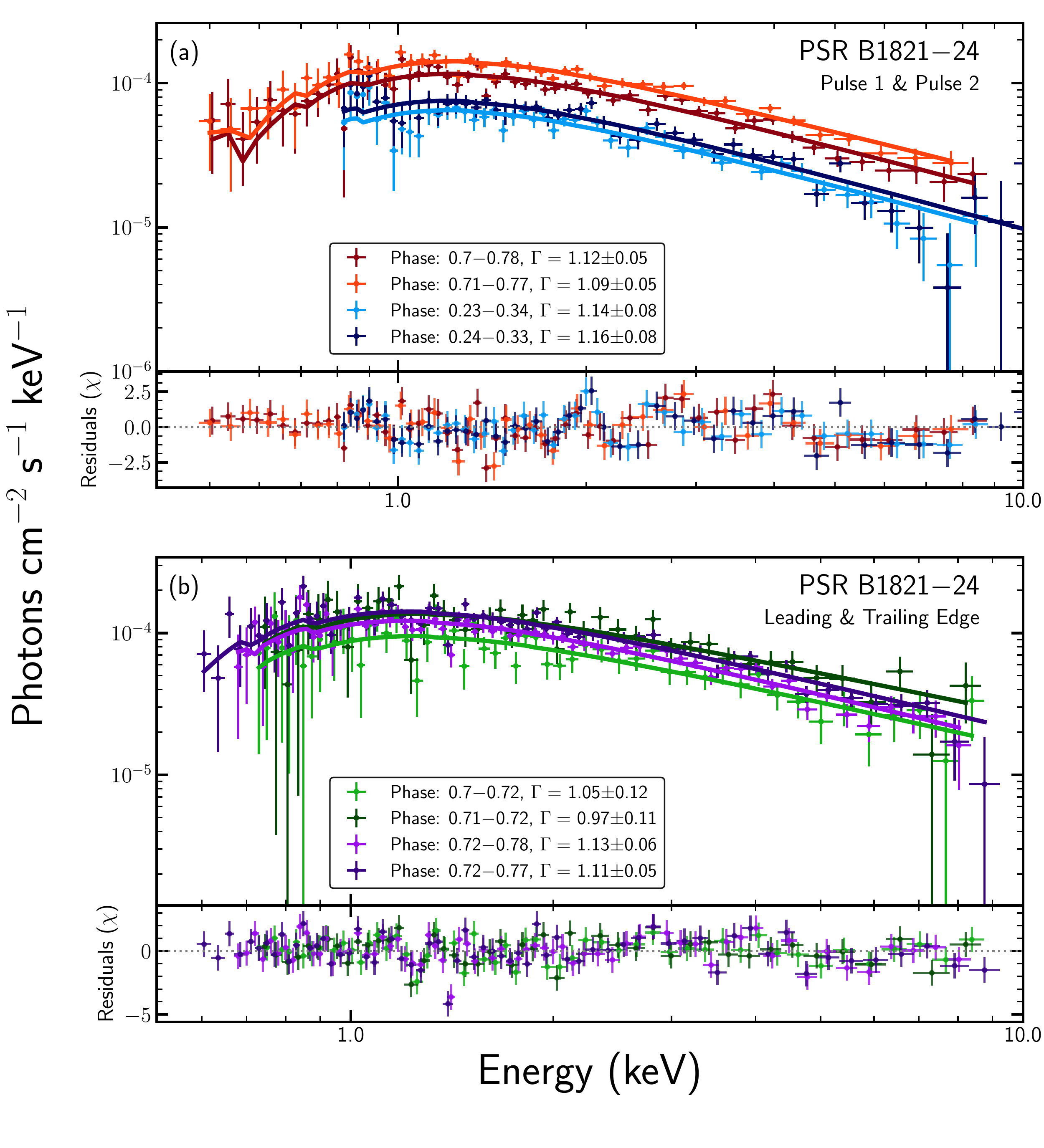}
    \caption{Emission spectra of two PSR~B1821$-$24 pulse components (a) and leading and trailing edges of P1 (b). We fit a single absorbed power-law to each of the phase selections. Below each set of spectra are the residuals of the best fit power-law in units of $\chi$.}
    \label{fig:1821_spectra}
\end{figure}

We then compare the emission spectra of each pulse component and the edges of P1. For each extracted spectrum, we fix the column density at $N_{\rm{H}}=0.38\times10^{22}$~cm$^{-2}$, the best-fit value from the pulsed emission. Figure~\ref{fig:1821_spectra} shows the emission spectra for these phase selections. Unlike the results of PSR~B1937$+$21, we find no significant difference in photon index between the spectrum of the two pulses, a result consistent with \textit{NuSTAR} analysis \citep{Gotthelf17}. The fit photon index of the spectra from the leading and trailing edge of P1 are also consistent within the 90\% confidence intervals. 

\begin{figure}
    \centering
    \includegraphics[width=\linewidth]{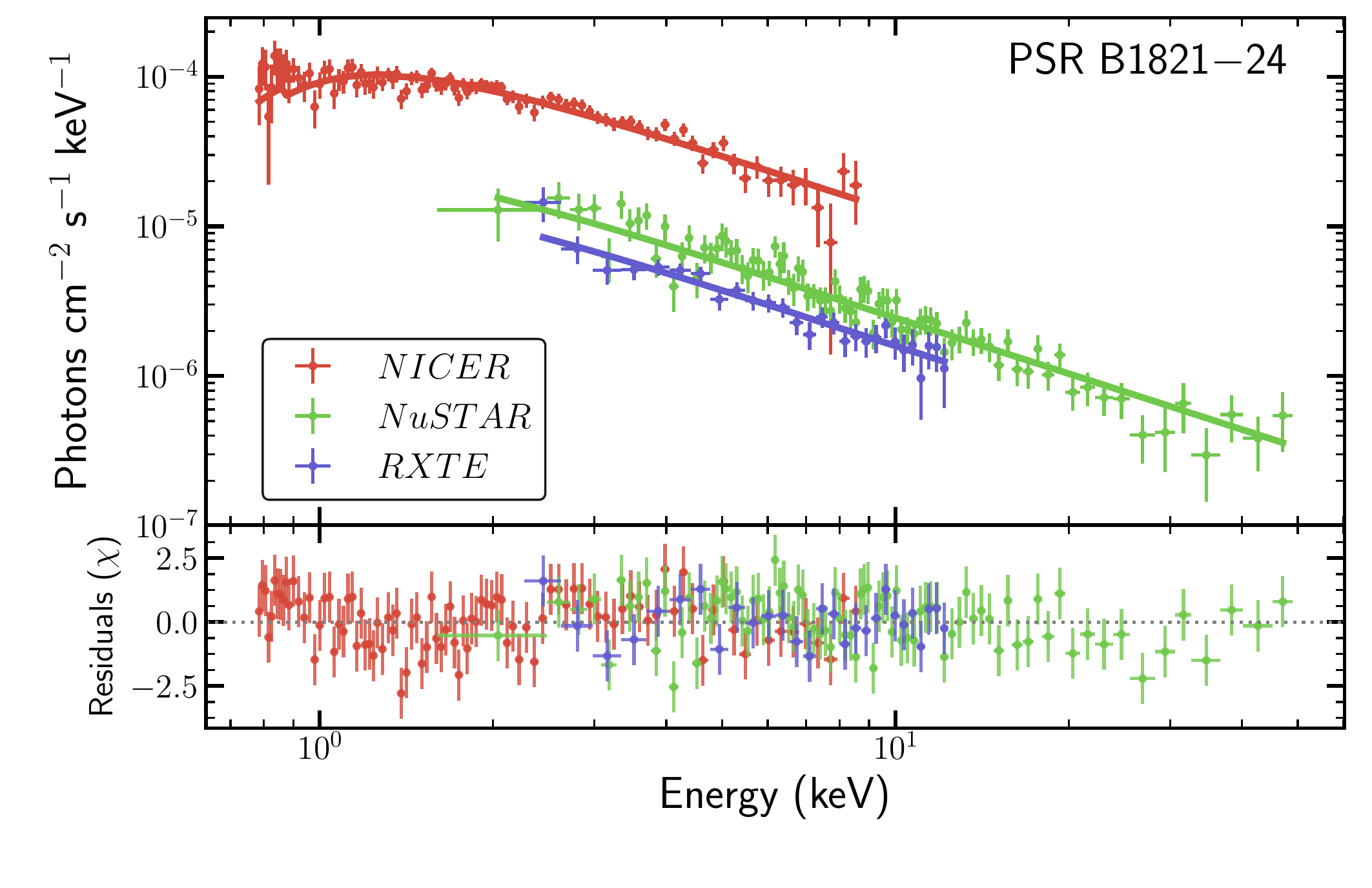}
    \caption{Top: Emission spectra of PSR~B1821-24 primary pulse component fit jointly with data from \textit{NuSTAR} and \textit{RXTE}. Bottom: Residuals of combined fit for each spectra.}
    \label{fig:1821_joint_spectra}
\end{figure}

We combine \textit{NICER} observations with \textit{NuSTAR} and \textit{RXTE} extracted spectra from \cite{Gotthelf17} to model the joint spectrum shown in Figure~\ref{fig:1821_joint_spectra}. Both  the \textit{NuSTAR} and \textit{RXTE} emission spectra use the scaled off-pulse region as the background. We find the best-fit single absorbed power-law has a photon index $\Gamma = 1.24 \pm 0.05$ and a hydrogen absorbing column density $N_{\rm{H}}=0.50\pm0.06 \times 10^{22}$~cm$^{-2}$ with a $\chi^2_{\nu} = 0.95$ for 213 dof.

\section{PSR~J0218+4232} \label{0218_section}

PSR~J0218+4232 was first confirmed as an MSP by \cite{Navarro95}, who showed it is in a two-day orbit with a white dwarf companion. It has a 2.3 ms period and a spin down luminosity of $\Dot{E} = 2.4 \times 10^{36}$~erg\,s$^{-1}$. This pulsar was detected in soft X-rays by \textit{ROSAT} HRI and PSPC \citep{Verbunt96_ROSAT, Kuiper98}. These observations revealed that two-thirds of the emission between 0.1--2.4~keV is non-pulsed.

PSR~J0218$+$4232 was also detected with the Energetic Gamma Ray Experiment Telescope (\textit{EGRET}) onboard the Compton Gamma Ray Observatory (\textit{CGRO}) \citep{Kuiper00}. \textit{BeppoSax} \citep{Mineo00}, \textit{NuSTAR} \citep{Gotthelf17}, and  \textit{XMM-Newton} \citep{Webb04} have continued to observe the X-ray emission of PSR~J0218$+$4232. As compared to PSR~B1821$-$24 and PSR~B1937$+$21, the functional form of the PSR~J0218$+$4232 spectra is less certain. The best fit model has been determined to be an absorbed power-law within the energy range 2--10~keV by \textit{BeppoSax} \citep{Mineo00} and \textit{NuSTAR} \citep{Gotthelf17}. Additionally, for the energy range of 0.2--10~keV, \textit{XMM-Newton} data suggests that the model of best fit is an absorbed power-law and a blackbody \citep{Webb04}. The variation in best fit models suggests that there may be a thermal component detected only at lower energies, possibly between 0.6--2.0~keV. \textit{NICER} has observed this source for $\sim1190$~ks between 2017-06-26 and 2019-06-23 in 282 ObsIDs. 

\subsection{Pulse profiles}

\begin{table*}
    \begin{lrbox}{\leftbox}
    \begin{tabular}{cc}
\toprule
Energy range (keV) & P1/P2\\

\midrule
0.5--2.0  & 1.6$\pm$0.1\\
2.0--4.0  & 1.5$\pm$0.1\\
4.0--12.0 & 1.7$\pm$0.3\\
\bottomrule
\end{tabular}
%updated with new phases
    \end{lrbox}

    \begin{lrbox}{\rightbox}
    \begin{tabular}{lllccccrr}
\toprule
           Region &   Phase Range &       $\Gamma$ &  $\chi^2_{\nu}$ &  Degrees of &  Number of \\
& & & & Freedom & Photons\\ 
\midrule
 \textbf{Pulse 1} &  0.28 -- 0.54 &  0.99$\pm$0.04 &            0.88 &       126.0 &     196111 \\
                  &  0.34 -- 0.48 &  0.98$\pm$0.04 &            0.89 &        80.0 &     112050 \\
     Leading Edge &  0.28 -- 0.41 &  1.02$\pm$0.06 &            0.82 &        88.0 &      97242 \\
\midrule
                  &  0.34 -- 0.41 &  0.98$\pm$0.06 &            0.93 &        52.0 &      55150 \\
    Trailing Edge &  0.41 -- 0.54 &  0.96$\pm$0.06 &            1.30 &        83.0 &      98869 \\
                  &  0.41 -- 0.48 &  0.95$\pm$0.06 &            0.99 &        49.0 &      56900 \\
 \textbf{Pulse 2} &  0.79 -- 1.03 &  0.92$\pm$0.07 &            0.99 &       104.0 &     171697 \\
                  &  0.85 -- 0.97 &   0.9$\pm$0.07 &            0.99 &        61.0 &      89814 \\
     Leading Edge &  0.79 -- 0.91 &   0.78$\pm$0.1 &            1.19 &        67.0 &      85255 \\
                  &  0.85 -- 0.91 &  0.76$\pm$0.09 &            1.16 &        37.0 &      44867 \\
    Trailing Edge &  0.91 -- 1.03 &  1.03$\pm$0.08 &            0.92 &        67.0 &      86442 \\
                  &  0.91 -- 0.97 &  0.94$\pm$0.09 &            0.92 &        38.0 &      44947 \\
\bottomrule
\end{tabular}

    \end{lrbox}

    \centering
    \makebox[0pt]{%
        \begin{minipage}[b]{\wd\leftbox} % A minipage that covers half the page 
            \centering
            \caption{Evolution of PSR~J0218$+$4232 P1/P2 peak ratio with respect to energy.} \label{tab:0218_PR}
            \usebox{\leftbox}
        \end{minipage}\quad
        \begin{minipage}[b]{\wd\rightbox}
            \centering
            \caption{Best-fit model parameters for the single absorbed power-law $+$ Gaussian PSR~J0218$+$4232 emission spectra at different phase selections. Since the number of photons varies between phase selections, we use different binning when extracting the spectra, resulting in different degrees of freedom.} \label{tab:0218_models}
            \usebox{\rightbox}
        \end{minipage}%
    }
\end{table*}

\begin{figure}
    \centering
    \includegraphics[width=\linewidth]{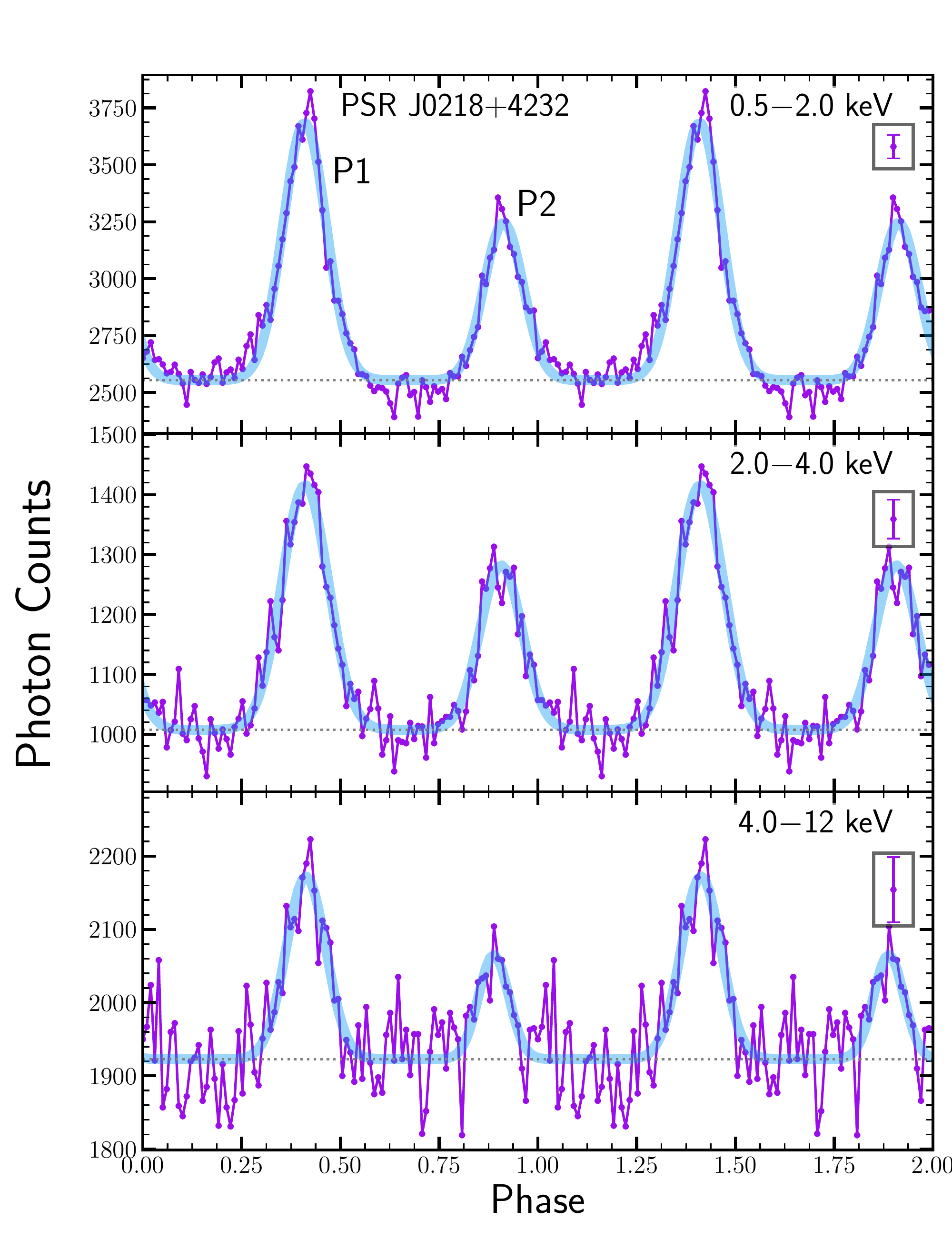}
    \caption{Phase folded pulse profile for PSR~J0218$+$4232 with 100 phase bins. We fit Gaussians to each component to estimate its center and width. The pulse components are broader as compared to the other observed pulsars. The boxed point shows the characteristic error bar for each profile. The gray dotted line represents the vertical offset when fitting the double Gaussian model.}
    \label{fig:0218_profiles}
\end{figure}

The phase folded pulse profile, shown in Figure~\ref{fig:0218_profiles}, is much broader than those of the other two pulsars presented in this paper. There is evidence for an emission bridge connecting the two pulses. Unlike the pulses of the other two pulsars, the pulses of PSR~J0218$+$4232 are best fit by Gaussians rather than Lorentzians. 

\begin{figure}
    \centering
    \includegraphics[width=\linewidth]{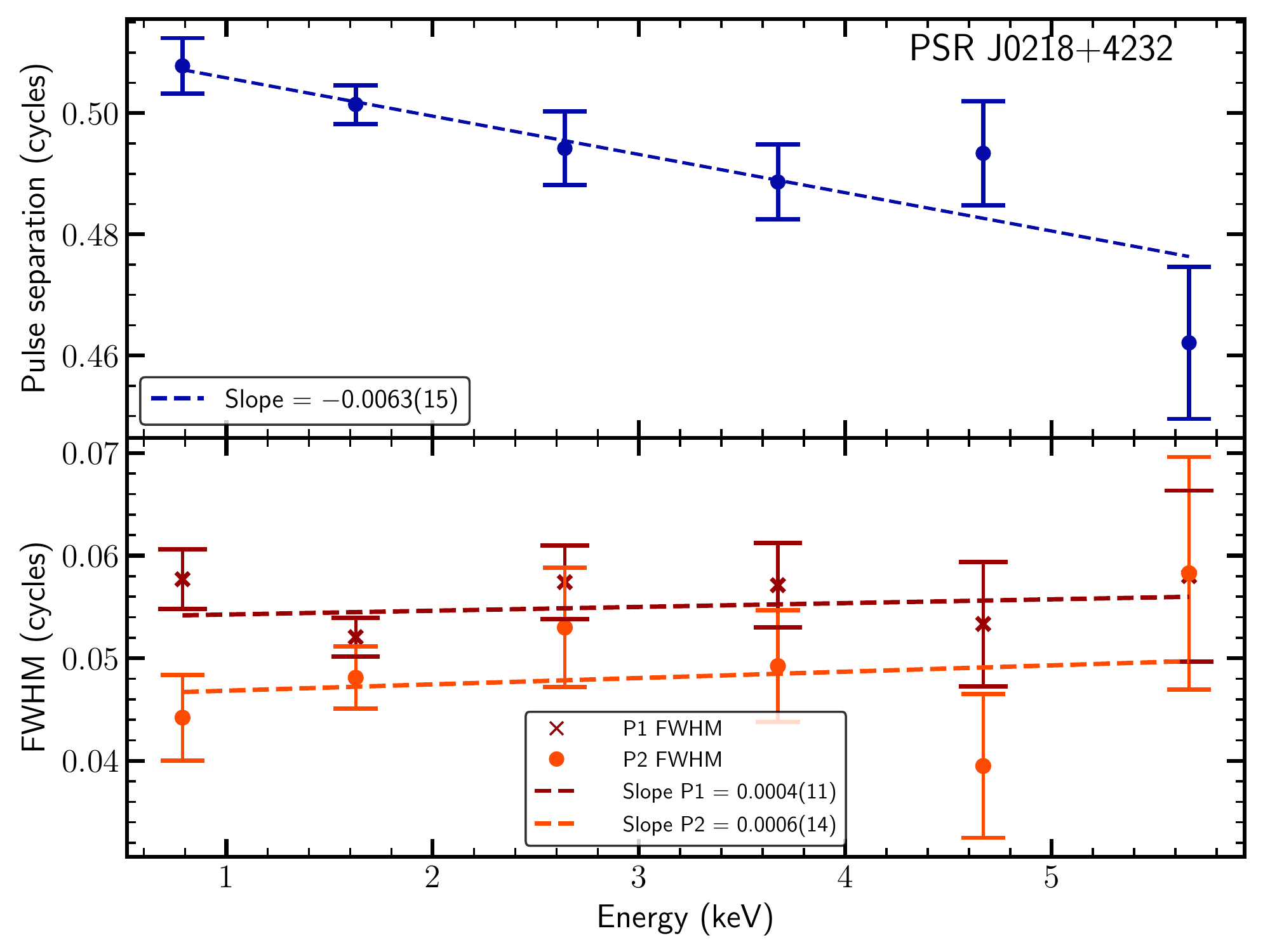}
    \caption{Evolution of pulse separation and FWHM as a function of energy for PSR J0218$+$4232}
    \label{fig:0218_PP}
\end{figure}

As before, we utilize a three-step fitting process. We first fit two Gaussians to the pulse profile at six energy subsets between 0.2--6.2~keV. We then monitor the evolution of pulse separation with slopes $m_{\rm{sep}}$, $m_{\rm{FWHM,P1}}$, and $m_{\rm{FWHM,P2}}$. Figure~\ref{fig:0218_PP} shows the results of our fitting procedure.

We find a significant decrease in pulse separation at the 3.7$\sigma$ level with higher energies. A detailed discussion of these results is offered in section \ref{discussion_conclusion_section}. This is the first detection of energy dependent pulse separation in PSR J0218$+$4232 X-ray observations. 

We find that the slopes corresponding to each pulse width, $m_{\rm{FWHM,P1}}$ and $m_{\rm{FWHM,P2}}$, are consistent with zero for both P1 and P2. We find the $95\%$ upper limit for width evolution is 0.00071 cycles/keV for |$m_{\rm{FWHM,P1}}$| and 0.0012 cycles/keV for |$m_{\rm{FWHM,P2}}$|.

Table \ref{tab:0218_PR} shows our measurement of the ratio P1/P2 is consistent with no change with energy, although a higher signal-to-noise measurement could reveal a trend.

\setlength{\tabcolsep}{6pt}
\begin{table}
\centering
    \caption{Pulse separation and width corresponding to Figure \ref{fig:0218_PP}. For each energy range, we use the pulsed emission spectra from \S\ref{0218_spectra_section} to find the median energy. We find the slope of the pulse separation is significant at the 3.7 $\sigma$ level.}
    \label{tab:0218_PP}
    \begin{tabular}{ccccccc}
\toprule
 Pulse & FWHM & FWHM & Energy & Energy \\
 Separation & of P1 &of P2 & band&Median\\
 (cycles) &(cycles)&(cycles)&(keV)&(keV)\\
\midrule
0.508(4) & 0.058(3) & 0.044(4) & $0.2-1.2$ & $0.79$\\
0.501(3) & 0.052(2) &  0.048(3) & $1.2-2.2$ &  $1.63$ \\
0.494(6) & 0.057(3) &  0.053(6) & $2.2-3.2$ & $2.64$\\
0.489(6) & 0.057(4) & 0.049(5) &  $3.2-4.2$ & $3.67$\\
0.493(9) & 0.053(6) & 0.039(7) & $4.2-5.2$ & $4.67$\\
0.462(12) & 0.058(5) & 0.058(11) & $5.2-6.2$ & $5.67$\\
\bottomrule
\end{tabular}
\end{table}

\subsection{Spectroscopy} \label{0218_spectra_section}

\begin{figure}
    \centering
    \includegraphics[width=\linewidth]{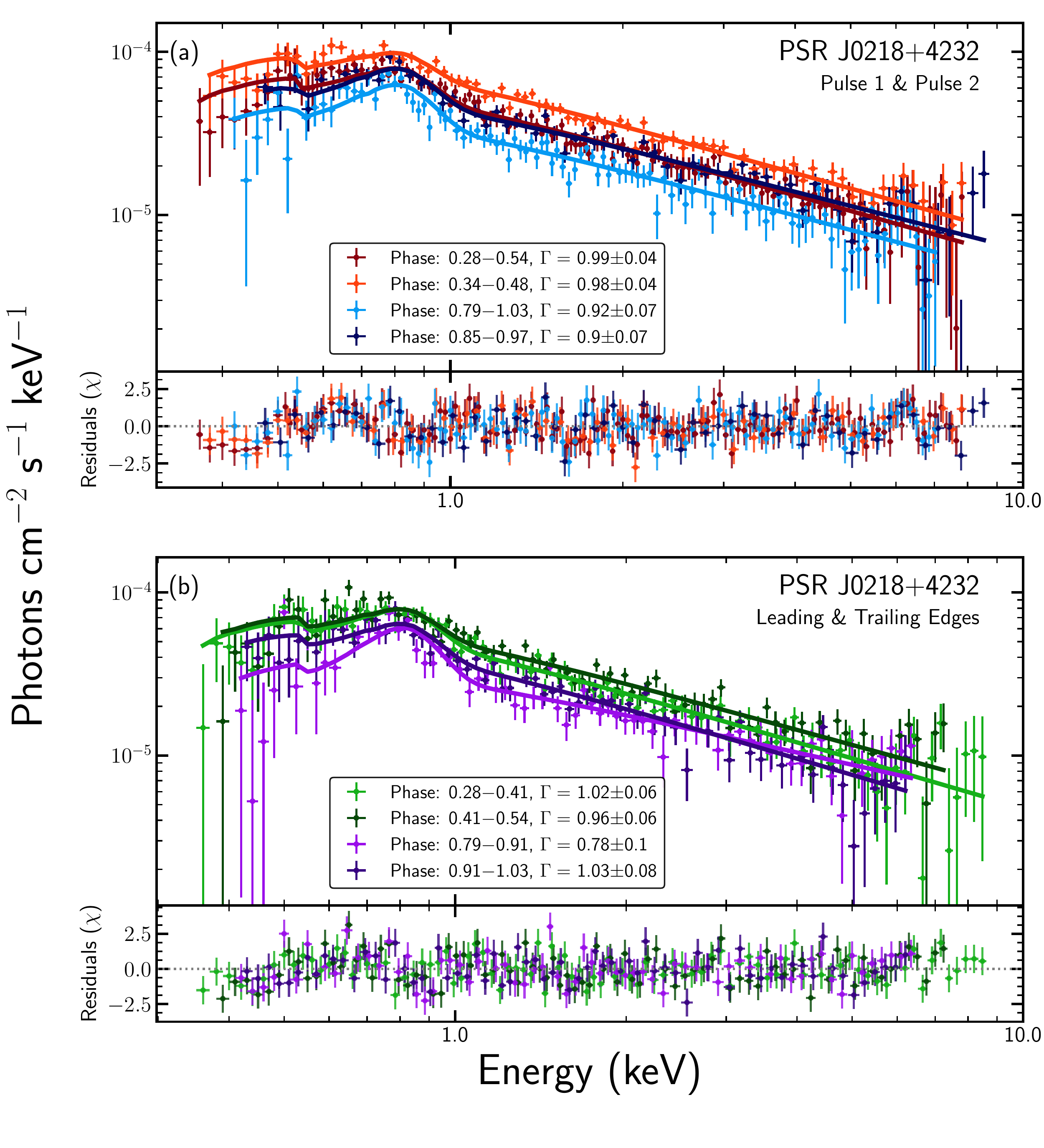}
    \caption{Emission spectra of two PSR~J0218$+$4232 pulse components (a) and leading and trailing edges of P1 and P2 (b). We fit single absorbed power-laws with a Gaussian emission line to each phase selection. Below each set of spectra are the residuals of the model in units of $\chi$.}
    \label{fig:0218_spectra}
\end{figure}

Since the pulsed fraction is only $\sim 65\%$ for this source \citep{Zavlin07}, we can not rely on the off-pulse region for our selecting the phase boundaries when extracting our spectra. Instead, we use the Gaussian fitted parameters from the previous section to determine the centers and widths of the pulses. We first extract the spectrum between phases $0.28 \leq \phi \leq 0.54$ for P1 and $0.79 \leq 1.03$ for P2. Since the phase ranges are wider for this pulsar, and the exposure time is greater, we group the extracted spectra with minimum 2500 counts/bin while maintaining similar energy resolution per bin as the other pulsars. 

Rather than defining a narrow off-pulse region to extract a background spectrum, we use a space-weather background generated based on the geomagnetic index $K_p$ and cutoff rigidity {\tt COR\_SAX}. We first fit a single absorbed power-law to the combined P1$+$P2 pulsed emission. The HI column density is fixed at $N_{H}=6.75\times10^{20}$~cm$^{-2}$, the value from \cite{HI4PICollab16}. The best fit has a photon index $\Gamma=1.19$ with a high $\chi^2_{\nu}=1.93$ for 84 dof. Since this background is based on the GTIs used in the extracted spectra, our background subtracted spectra still shows residual features at $<1.0$~keV, possibly due to ionized oxygen emission. This feature, likely an emission line originating in the Solar wind or local hot bubble, has been observed in \textit{NICER} spectra before \citep{Ray19}.

In order to improve upon the fit, we add an additional Gaussian component. The best fit {\tt tbabs*(powerlaw+gaussian)} model has photon index $\Gamma=1.05\pm0.05$, line center $l_E=0.80\pm0.03$~keV, and line width $\sigma_E = 0.11\pm0.03$~keV resulting in a $\chi^2_{\nu}=0.88$ for 82 dof.

We then fit the spectrum extracted from narrow phase regions, including both pulses and their leading/trailing edges. For each model, we freeze the $l_E$ and $\sigma_E$ at the values listed above because the foreground emission component is not expected to vary significantly between phase selections. Figure~\ref{fig:0218_spectra} plots the emission spectra with the {\tt tbabs*(powerlaw+gaussian)} model for each phase selection. Table \ref{tab:0218_models} gives the model fits for each phase range. While our results seem to suggest that the P1 photon index is higher than that of P2, we find that whether or not the confidence intervals overlap depends on the phase selections used. 

\begin{figure}
    \centering
    \includegraphics[width=\linewidth]{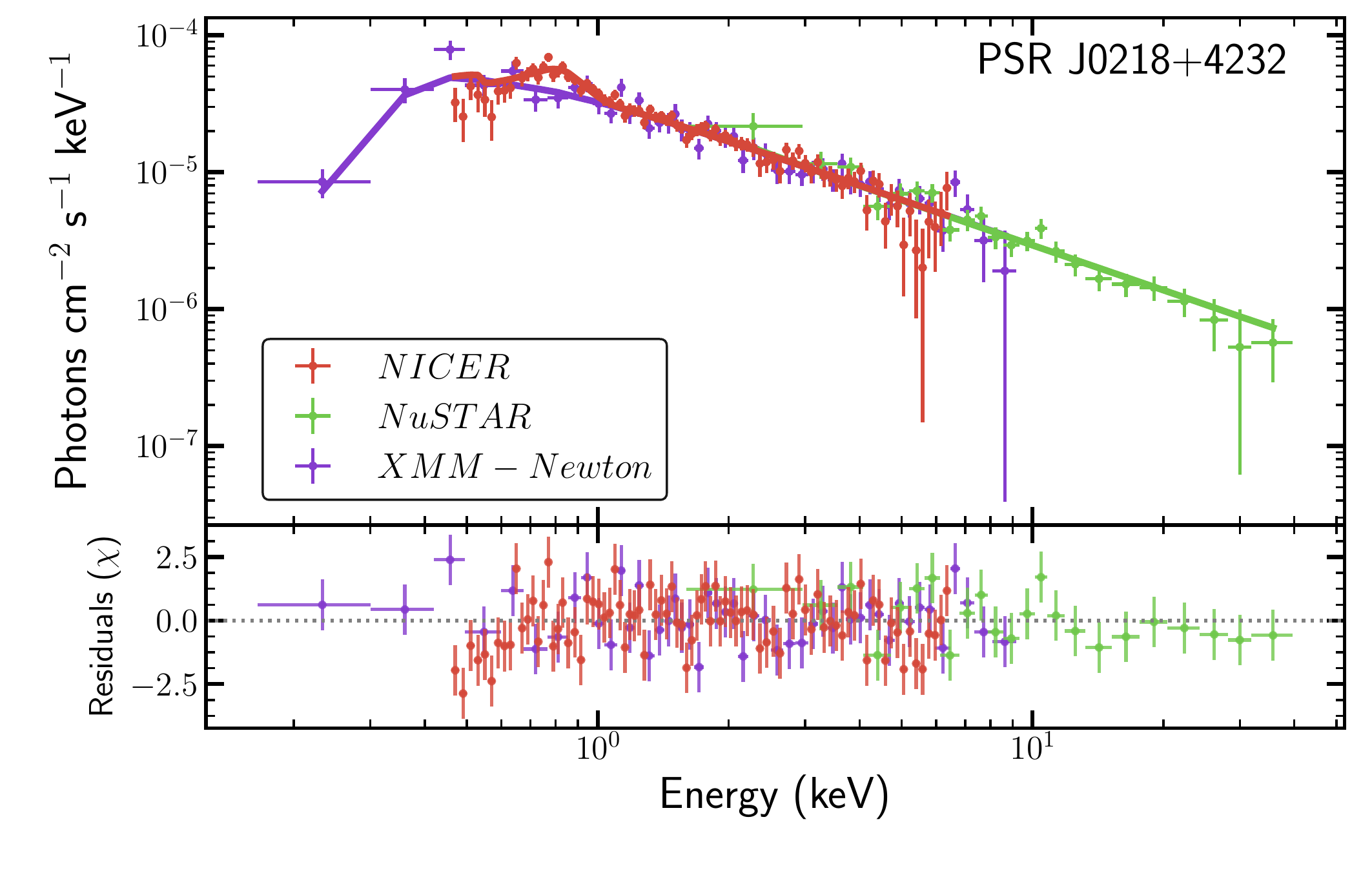}
    \caption{Top: Emission spectra of PSR~J0218$+$4232 extracted from both pulses fit jointly with data from \textit{NuSTAR} and \textit{XMM-Newton}. Bottom: Residuals of the joint model fit for each spectra.}
    \label{fig:0218_joint_spectra}
\end{figure}

Finally, we fit the \textit{NICER} extracted spectrum alongside \textit{NuSTAR} and \textit{XMM-Newton} observations from \cite{Gotthelf17}. The \textit{XMM-Newton} spectra was originally presented in \citep{Webb04} and uses a off-source image for the background spectra. The \textit{NuSTAR} spectra uses a phase averaged background spectra. Again, we fix the column density $N_{\rm{H}} = 6.75\times10^{20}$~cm$^{-2}$. The best fit single-absorbed power-law has a photon index $\Gamma=1.18\pm0.03$ with a $\chi^2_{\nu}=1.47$ for 154 dof. We also fit the {\tt tbabs*(powerlaw+Gaussian)} model, freezing the Gaussian component parameters at zero for the \textit{NuSTAR} and \textit{XMM-Newton} data. The best-fit model has a photon index $\Gamma=1.10\pm0.04$ with a $\chi^2_{\nu}=1.06$ for 151 dof. Figure~\ref{fig:0218_joint_spectra} plots the simultaneous fit of these three telescopes.

\section{Discussion \& Conclusions}
\label{discussion_conclusion_section}
We present two years of \textit{NICER} soft X-ray observations of three energetic MSPs: PSRs~B1937$+$21, B1821$-$24, and J0218$+$4232. We use the high precision timing measurements to track pulse profile evolution at different energies and model phase-resolved emission spectra.

Table \ref{tab:slopes} summarizes the results of the analysis of profile evolution with energy for the three pulsars. We find that the decrease of pulse separation is significant at $>3.7\sigma$ for PSR~J0218$+$4232 in the \textit{NICER} energy range. This is the first evidence for the evolution of pulse profile separation in the soft X-ray regime for this pulsar.

While pulse profile morphology has been studied extensively in radio observations \citep[e.g.,][]{Rankin83_morphtax}, few pulsars are bright enough for detailed measurements at higher energies. Observations of the Vela pulsar (PSR~J0835$-$4510) with the Large Area Telescope on \textit{Fermi} have revealed constant pulse separation between pulse 1 and pulse 2 and an increasing separation between pulse 1 and pulse 3 as a function of energy. The widths of all three pulses were found to decrease \citep{Abdo10}. Observations of the Crab Pulsar have shown similar results of no phase shift with energy and decreasing pulse width \citep{Abdo10_crab}.

The energy dependence of pulse separation for PSR~J0218$+$4232 suggests that regions characterized by slightly higher particle energies, larger local magnetic fields, or larger pitch angles map to slightly different phases of the observed light curve than the lower-energy radiation. Our viewing angle could be such that
P1 and P2 could originate near the same side of the CS associated with one magnetic pole. In this case, the higher-energy emission originates slightly farther from the CS center compared to the lower-energy emission. The magnetic field lines going into the CS map back to the return current region on one polar cap \citep{Timokhin13, Harding18}. On the other hand, if P1 and P2 originate from opposite poles, then they come from opposite sides of the respective CSs, with higher-energy particles again originating farther from the CS center for the two-pole model to be consistent with the peaks' frequency-dependent shift (see e.g. Figure~10 in \citealt{Bai10}). 

We find no evidence for energy-dependent pulse separation in PSR~B1937$+$21 and PSR~B1821$-$24 and place constraints on the X-ray pulse separation evolution. A higher instrumental sensitivity could reveal energy-dependent separations for these pulsars.

Our results for FWHM evolution suggest that pulse width increases in PSR B1937$+$21 and decreases in PSR B1821$-$24 at confidence levels ranging from 0.8~$\sigma$ to 2.6~$\sigma$, while it remains constant across the \textit{NICER} energy regime for PSR J0218$+$4232.

\setlength{\tabcolsep}{5pt}
\begin{table}
\centering
    \caption{Slopes of the linear regression of pulse separation and FWHM with respect to energy.}
    \label{tab:slopes}
    \begin{tabular}{ccccc}
\toprule
PSR & Pulse Separation & FWHM of P1 &  FWHM of P2\\
\midrule
B1937$+$21 & 0.001(1)  & 0.00012(10) & 0.0026(7)\\
B1821$-$24 & 0.00028(39) & $-$0.00012(8) & $-$0.0007(3)\\
J0218$+$4232 & $-$0.0063(15) & 0.0004(11) & 0.0006(14)\\
\bottomrule
\end{tabular}

    \footnotesize
    \emph{\newline The errors correspond to the 1-$\sigma$ Gaussian standard deviations.}
\end{table}

Spectral analyses are consistent with the hard power-law emission spectra expected for these three pulsars. With over 1000~ks of exposure for PSRs~B1937$+$21 and J0218$+$4232 and over 700~ks of exposure for PSR~B1821$-$24, \textit{NICER} observations contribute to the wealth of X-ray observations used to model the emission spectra of these pulsars. We provide updated model parameters of the single absorbed power-law for each pulsar and find no evidence of additional black body components for PSRs~B1937$+$21 and B1821$-$24. We observe an additional emission line feature at $<1.0$~keV for PSR J0218$+$4232 unfit by the instrument response. This limitation prevents accurate evaluation of black body components for this pulsar. 

For PSR~B1937$+$21, we observe different emission spectra photon indices at $>2\sigma$ level. Observation of different photon indices for different pulse profile peaks suggests that the particle spectrum varies between phases. This may be due to slightly different local conditions such as the magnetic field strengths and pitch angles, or different pair injection spectra, which influence the dynamics and radiation by these particles. We do not observe this behavior for PSR~B1821$-$24, suggesting a similar emission origin or local conditions for each peak.

Though the majority of equatorial CS models pertain to gamma-ray emission \citep[e.g.][]{Cerutti16, Kalapotharakos14}, we note that the pulsars presented here have pulse features that are aligned across radio, X-ray, and gamma-ray observations. In the case of PSR~B1937$+$21, a close phase alignment between radio giant pulses and X-ray emission has been observed \citep{Cusumano03}, motivating \textit{NICER} correlation searches that will be reported in a separate paper. Profiles that are phase-aligned across wavebands may point to (nearly) overlapping spatial emission origins for the different bands. The closer the overlap in altitude and extent of these regions, the closer the phase-alignment across bands. In earlier gap models, such as slot gap \citep{Muslimov04} and outer gap \citep{Romani95}, the special relativistic effects of aberration and time-of-flight delay plus the magnetic field geometry \citep{Dyks04} affect photons of different energies in the same way to produce caustics in the emission sky maps. In more modern global magnetosphere models, caustics form due to stagnation of emission directions as a result of the magnetic field geometry in the CS \citep{Bai10}. While the origin of the caustics is qualitatively different in these models, the argument of similar spatial origin of photons seen at similar observational phases continues to hold.

The present X-ray observations thus probe both the emission geometry and  spatial properties of the plasma in the CS, which will be augmented as future models will attempt to produce the correct peak phases in many different wavebands.

\section*{Acknowledgements}
This work was supported by NASA through the \textit{NICER} mission and the Astrophysics Explorers Program. DR, LL, and ZG acknowledge support from the NANOGrav Physics Frontiers Center (NSF award number 1430284) and from Haverford College through the Louis Green Fund. ES acknowledges support from the Marian E. Koshland Integrated Natural Sciences Center Summer Scholar fund at Haverford College. RML acknowledges the support of NASA through Hubble Fellowship Program grant HST-HF2-51440.001.

The authors thank K. Kalapotharakos for his insight on interpretation of our results, and M. Corcoran for his assistance in generating the space-weather model background. We also thank J. Cammisa for his input on data acquisition and handling. 

This research has made use of NASA's Astrophysics Data System Bibliographic Services (ADS) and the arXiv.

\software{
    \texttt{HEAsoft} \citep[v6.26.1][]{Nasa14}, 
    \texttt{PINT} (\url{https://github.com/nanograv/pint}), 
    \texttt{XSELECT} (v2.4),
    \texttt{XSPEC} \citep[v12.1.0][]{Arnaud96}, 
    \texttt{NICERSoft} (\url{https://github.com/paulray/NICERsoft}).}
\facility{\textit{NICER}}

%%%%%%%%%%%%%%%%%%%%%%%%%%%%%%%%%%%%%%%%%%%%%%%%%%



\bibliographystyle{aasjournal}
\bibliography{NicerBib} 

%%%%%%%%%%%%%%%%%%%%%%%%%%%%%%%%%%%%%%%%%%%%%%%%%%

%%%%%%%%%%%%%%%%% APPENDICES %%%%%%%%%%%%%%%%%%%%%

\appendix
\section{Different Profile Fitting Methods} \label{appendixFitting}

\begin{figure}
    \centering
    \includegraphics[width=\linewidth]{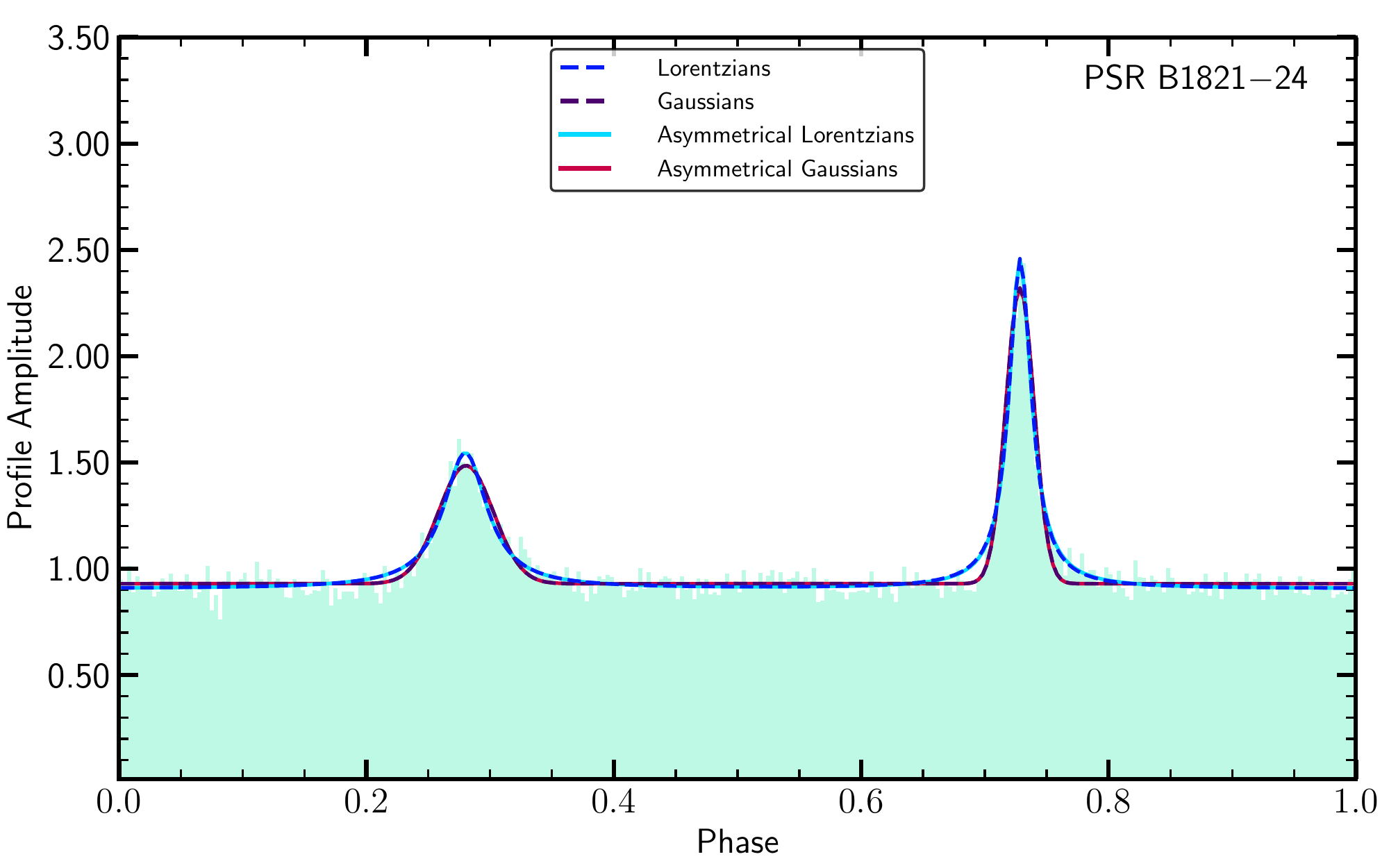}
    \caption{Fitting the PSR~B1821$-$24 pulse profile against different functional forms in the 1.0--2.0~keV energy band.}
    \label{fig:FIT}
\end{figure}

We used a variety of functional models to fit the pulse profile components in our study of profile evolution. The simplest model is a symmetrical Gaussian for each component \citep{Kramer94}. We use this model as a basis to compare other functional forms. By eye, the pulse components of PSRs~B1937$+$21 and B1821$-$24 appear slightly asymmetrical. Profile modeling of \textit{Chandra} \citep{Ng14} and \textit{Fermi} \citep{Abdo13} observations used a sum of two asymmetrical Lorentzians. Some sources suggest that the asymmetrical pulse component is due to giant pulse emission at the trailing edge \citep{Romani01}, motivating the implementation of a triple Gaussian model. Figure~\ref{fig:FIT} plots each of the aforementioned functional forms for PSR~B1937$+$21's pulse components. The triple-Gaussian model was discarded for being overly dependent on initial conditions and thus unable to give a repeatable fit. Repeating the fitting procedure with different initial conditions for the third Gaussian component led to different results. Only the symmetrical and asymmetrical Gaussians and Lorentzians were compared. Because the asymmetrical functions were made nearly symmetrical by the parameter optimization algorithm, a $\chi^2$ analysis proved the addition of the skewness parameters to be inefficient. Over the five energy ranges for PSRs~B1937$+$21 and B1821$-$24, symmetrical Lorentzians outperformed symmetrical Gaussians with a significantly smaller $\chi^2$ value and were therefore chosen for our pulse profile analysis. For the PSR~J0218$+$4232 pulse profile, the Gaussian fits to each pulse are a better fit and were used in the analysis of this pulsar. 

For each of these models, we compared the measurements of pulse component evolution with energy. We found that results for each pulsar are consistent across choice of model and phase binning. Table \ref{tab:phase_bin} shows no significant change in the slopes of profile feature evolution with energy for PSR J0218$+$4232 when adding more phase bins. Our results are thus independent of phase binning.

\begin{table}
    \centering
    \begin{tabular}{cccc}
\toprule
\vtop{\hbox{\strut Number of }\hbox{\strut phase bins}} & $m_{\rm{sep}}$ & $m_{\rm{FWHM,P1}}$ & $m_{\rm{FWHM,P2}}$\\
\midrule
50 &  -0.0066(15) & 0.0004(10) & 0.0006(13)\\
100 & -0.0063(15) & 0.0004(11) & 0.0006(14)\\
300 & -0.0063(14) & 0.0002(13) & 0.0005(15)\\

\bottomrule
\end{tabular}
    \caption{Comparison of profile feature slopes for PSR J0218$+$4232 using different numbers of phase bins.  }
    \label{tab:phase_bin}
\end{table}

%%%%%%%%%%%%%%%%%%%%%%%%%%%%%%%%%%%%%%%%%%%%%%%%%%

% Don't change these lines
%\bsp	% typesetting comment
\label{lastpage}
\end{document}